\documentclass[12pt,preprint]{aastex}
\usepackage{pslatex}
\usepackage{amssymb}

\shorttitle{Fast shock acceleration}
\shortauthors{Malkov and Diamond}

\newcommand\app{{\it Astropart. Phys.,\ }}
\newcommand\pr{{\it Phys.~Rev.,\ }}
\newcommand\jetp{{Sov.~Phys.~JETP,\ }}
\newcommand\procr{{\it Proc.\ 15th Int. Cosmic Ray Conf.\ }}
\newcommand\pfl{{\it Phys.~Fluids,\ }}
\newcommand\pop{{\it Phys.~Plasmas,\ }}
\newcommand\rpp{{\it Rep.~Progr.~Phys.,\ }}
\newcommand\ass{{\it Astroph. Space Sci.,\ }}

\newcommand\etal{{\it et al.~}}
\newcommand\eg{{\it e.g.,~}}
\newcommand\ie{{\it i.e.,~}}

\newcommand\jpg{{Journ. Phys. G.,\ }}

\newcommand\const{{\rm const\ }}

\begin{document}

\title{Nonlinear shock acceleration beyond the Bohm limit}

\author{M.A. Malkov and P.H. Diamond}

\email{mmalkov@ucsd.edu; phd@physics.ucsd.edu}

\affil{University of California at San Diego, La Jolla, California 92093-0319,
USA}

\begin{abstract}
We suggest a physical mechanism whereby the acceleration time of cosmic
rays by shock waves can be significantly reduced. This creates the
possibility of particle acceleration beyond the knee energy at $\sim10^{15}eV$.
The acceleration results from a nonlinear modification of the flow
ahead of the shock supported by particles already accelerated to the
knee momentum at $p\sim p_{*}$. The particles gain energy by bouncing
off converging magnetic irregularities frozen into the flow \emph{in
the shock precursor} and not so much by re-crossing the \emph{shock
itself}. The acceleration rate is thus determined by the gradient
of the flow velocity and turns out to be formally independent of the
particle mean free path (m.f.p.). The velocity gradient is, in turn,
set by the knee-particles at $p\sim p_{*}$ as having the dominant
contribution to the CR pressure. Since it is independent of the m.f.p.,
the acceleration rate of particles above the knee does not decrease
with energy, unlike in the linear acceleration regime. The reason
for the knee formation at $p\sim p_{*}$ is that particles with $p>p_{*}$
are effectively confined to the shock precursor only while they are
within limited domains in the momentum space, while other particles
fall into {}``loss-islands'', similar to the {}``loss-cone'' of
magnetic traps. This structure of the momentum space is due to the
character of the scattering magnetic irregularities. They are formed
by a train of shock waves that naturally emerge from unstably growing
and steepening magnetosonic waves or as a result of acoustic instability
of the CR precursor. These losses steepen the spectrum above the knee,
which also prevents the shock width from increasing with the maximum
particle energy.
\end{abstract}


\section{introduction}

The first order Fermi acceleration, also known as diffusive shock
acceleration (DSA) is regarded as the principal mechanism whereby
galactic cosmic rays (CRs) are produced. The physics of the energy
gain is very simple: particles get kicked sequentially by scattering
centers that are on the opposite side of the shock front and so approach
each other. Its modern version has been developed by \citet{krym77,axf77,bell78a,blaost78}
and others. Soon after that, however, it was realized that the mechanism
can only marginally \citep[if at all, ][]{lc83} account for the acceleration
of the galactic cosmic rays up to the knee energy at $10^{15}eV$
\citep[see also ][for more recent good discussions]{jones98,kirkd01}. Up to
the knee, the spectrum is nearly a perfect power-law so it is most
probably produced by a single mechanism. At the same time, the low
particle scattering rate makes the mechanism somewhat slow and the
particle confinement to the shock insufficient to accomplish this
task over the active life time of a typical galactic supernova remnant
(SNR) shock, which are considered to be the most probable sites of
CR acceleration. 

Not surprisingly, there have been many suggestions as to how to improve
the performance of the DSA. In one way or another they targeted the
most important and, at the same time, the most uncertain parameter
that determines the acceleration time, the particle diffusivity $\kappa$.
The acceleration rate can be written roughly as $\tau_{acc}^{-1}=U_{sh}^{2}/\kappa\left(p\right)$,
where $U_{sh}$ is the shock velocity. Therefore, the acceleration
rate \emph{decreases} with particle momentum $p$, since the particle
diffusion coefficient $\kappa$ usually \emph{grows} with it. In the
Bohm limit, which is reasonably optimistic, one can represent $\kappa$
as $\kappa=cr_{g}/3$, where $r_{g}$ is the particle gyroradius (substituted
in the last formula as a particle mean free path $\lambda$) and $c$
is the velocity of light. The condition $\lambda\sim r_{g}$ is justified
when the magnetic field perturbations that scatter accelerated particles
(and are driven by them at the same time) reach the level of the ambient
magnetic field, $\delta B\sim B_{0}$. Such a high fluctuation level
has been long considered as a firm upper limit \citep{McV82} and there
have been calculations indicating that the turbulence may saturate
at a somewhat lower level due to nonlinear processes  \citep[\eg][]{ab86}. 

It should be noted, however, that in cases of efficient particle acceleration,
\ie when a significant part of the shock ram pressure $\rho U_{sh}^{2}$
is converted to the pressure of accelerated particles $P_{c}$, there
is indeed sufficient free energy of accelerated particles which can
potentially be transformed to fluctuations with $\delta B\gg B_{0}$.
This would decrease $\lambda$ significantly below the particle gyroradius
(calculated by the unperturbed field $B_{0}$), so the acceleration
time would also decrease. From simple energetics principles, one can
estimate the maximum fluctuation energy as \citep{McV82}

\begin{equation}
\frac{\delta B}{B_{0}^{2}}^{2}\sim M_{A}\frac{P_{c}}{\rho U_{sh}^{2}}\label{eq:maxdb}\end{equation}
 where $M_{A}=U_{sh}/V_{A}\left(\textrm{B}_{\textrm{0}}\right)$ is
the shock Alfven Mach number. \citet{bell:luc} suggested that this
upper bound may indeed be reached corroborating this idea by the numerical
simulation \citep{lucb00}. The supporting simulation was rather limited
in the system size and particle energy range, which in turn, truncates
the wave spectrum. Therefore, the saturation effects predicted earlier
by \citet{McV82,ab86,be87}, have not been observed in the simulations
by \citet{lucb00}. In any case, eq.(\ref{eq:maxdb}) provides only
an upper bound to the actual wave amplitude which may be much lower
due to the saturation effects, not included in its derivation.

A different approach to the same goal of increasing the turbulent
component beyond the ambient field level have been pursued in the
recent papers by \citet{PZ03,ptus05}. These authors studied the DSA in the
presence of a Kolmogorov type MHD cascade initiated by the strong
MHD instability in the CR precursor. Also recently, \cite{bell04}
took up from the \cite{bell:luc} and \cite{lucb00} approaches and
replaced the time consuming particle simulations by providing an estimate
of the growth rate caused by the CR current at the shock. With the
current kept fixed, the waves indeed grow to a high level sufficient
to increase the maximum particle momentum by a factor of $10^{3}$.
It remains unclear, however, whether or not the CR trapping by the
self-generated turbulence and other wave-particle interactions
\citep[\eg][]{vlk84,ab86} saturates the instability at a lower level. Besides,
the alleged growth of $\delta B$ beyond $B_{0}$ clearly invalidates
treatments of the linear instability which are based on the $\delta B\ll B_{0}$
assumption. 

\cite{dm04} suggested an alternative picture in which a strong $\delta B$
may be generated as a result of a type of inverse cascade in $k$-space.
Strong field perturbations driven resonantly by already accelerated
particles are nonlinearly coupled to longer scales not only facilitating
the acceleration of higher energy particles but also providing an
ambient field for them as the longest scale field perturbation. Such
cascade can be driven by scattering of the Alfven waves on acoustic
waves driven, in turn, by the Drury instability of the CR precursor.
Another related route to long scale magnetic fluctuations is modulational
instability of the Alfven waves themselves. The condensation of magnetic
energy at long scales allows us to marginally preserve the weak turbulence
$\delta B<B_{0}$ requirement, where the role of $B_{0}$ is taken
by the longest scale part of the magnetic energy spectrum (longer
than the particle gyro-radius). This could possibly weaken the saturation
factors setting in at $\delta B\sim B_{0}$, which are noted above.

The above-mentioned works, while aimed at the explanation of acceleration
beyond the knee do not specifically address mechanisms that could
be responsible for the formation of the knee itself. An exception
is a paper by \cite{DruryKnee}, where the authors suggested that
the knee can appear as a result of an abrupt slowing down of the acceleration
at the sweep-up phase in combination with the Bell hypothesis about
the generation of the strong magnetic field. Some further interesting
ideas about the knee origin can be found in a recent paper by \cite{sveshn}.

In this paper we propose a different scenario of faster than Bohm
acceleration which is also intimately related to the knee phenomenon.
Odd as it sounds, this mechanism does not require super-Bohm magnetic
field fluctuations. The reason for that is that its rate does not
depend on the particle scattering mean free path $\lambda$. In the
next section we describe the mechanism by comparing it with both the
test particle and nonlinear DSA regimes under the Bohm diffusivity.
We analyze the conditions under which the proposed acceleration regime
is faster than the latter two. This sets the stage for Sec.\ref{sec:Particle-dynamics}
and \ref{sec:Particle-transport} where a preliminary study of particle
dynamics and transport in the CR precursor is presented. Section \ref{sec:Estimate-of-the maxmom}
deals with the estimates of the maximum energy that can be reached
in excess of the knee energy. In Sec.\ref{sec:Particle-spectrum-between}
an acceleration model that allows us to calculate the slope of the
spectrum between the knee and the maximum energy is presented. We
conclude with a summary and brief discussion.

\section{Acceleration Mechanism: a primer\label{sec:Acceleration-Mechanism}}

Perhaps the easiest way to understand why and when the suggested version
of the DSA becomes faster than the standard one, is to consider why
the latter is slow. For, we turn to the individual particle treatment
due to \cite{bell78a}. Upon completing one acceleration cycle, \ie
crossing and re-crossing the discontinuity, a particle gains momentum

\begin{equation}
\frac{\Delta p}{p}\sim\frac{\Delta U}{c}\label{eq:Dp}\end{equation}
 where $\Delta U$ is the relative velocity between the upstream and
downstream scattering centers $\Delta U=U_{1}-U_{2}\sim U_{1}$. Thus,
over the acceleration time (when the momentum gain $\Delta p\sim p$),
the number of cycles that the particle needs to make is $N_{cycl}\sim c/U_{1}\gg1$.
Apart from numerical factors and the differences between the upstream
and downstream residence time contributions, the particle acceleration
time can be estimated as 

\begin{equation}
\tau_{acc}\simeq\frac{\kappa\left(p\right)}{U_{1}^{2}}\sim\lambda c/U_{1}^{2}\sim\tau_{col}c^{2}/U_{1}^{2}\label{eq:tauacc}\end{equation}
where $\lambda$ and $\tau_{col}$ are the particle mean free path
(m.f.p.) and collision time, respectively. It is also useful to note
that the acceleration time is of the order of time needed to the fluid element
to cross the diffusion zone ahead of the shock, $\tau_{acc}\sim L_{dif}\left(p\right)/U_{1}$,
where $L_{dif}=\kappa\left(p\right)/U_{1}$ is the particle diffusion
length. Using eqs.(\ref{eq:Dp}) and (\ref{eq:tauacc}), one can write
the following relation between the acceleration time, the time needed
to complete one cycle and the collision time

\[
\tau_{acc}:\tau_{cycl}:\tau_{col}\sim\frac{c^{2}}{U_{1}^{2}}:\frac{c}{U_{1}}:1\]
Therefore, out of the $c^{2}/U_{1}^{2}$ wave-particle collisions,
needed to gain a momentum $\Delta p\sim p$, only $c/U_{1}$ are productive
in terms of the energy gain. Most of the collisions are wasted. 

\begin{figure}
\plotone{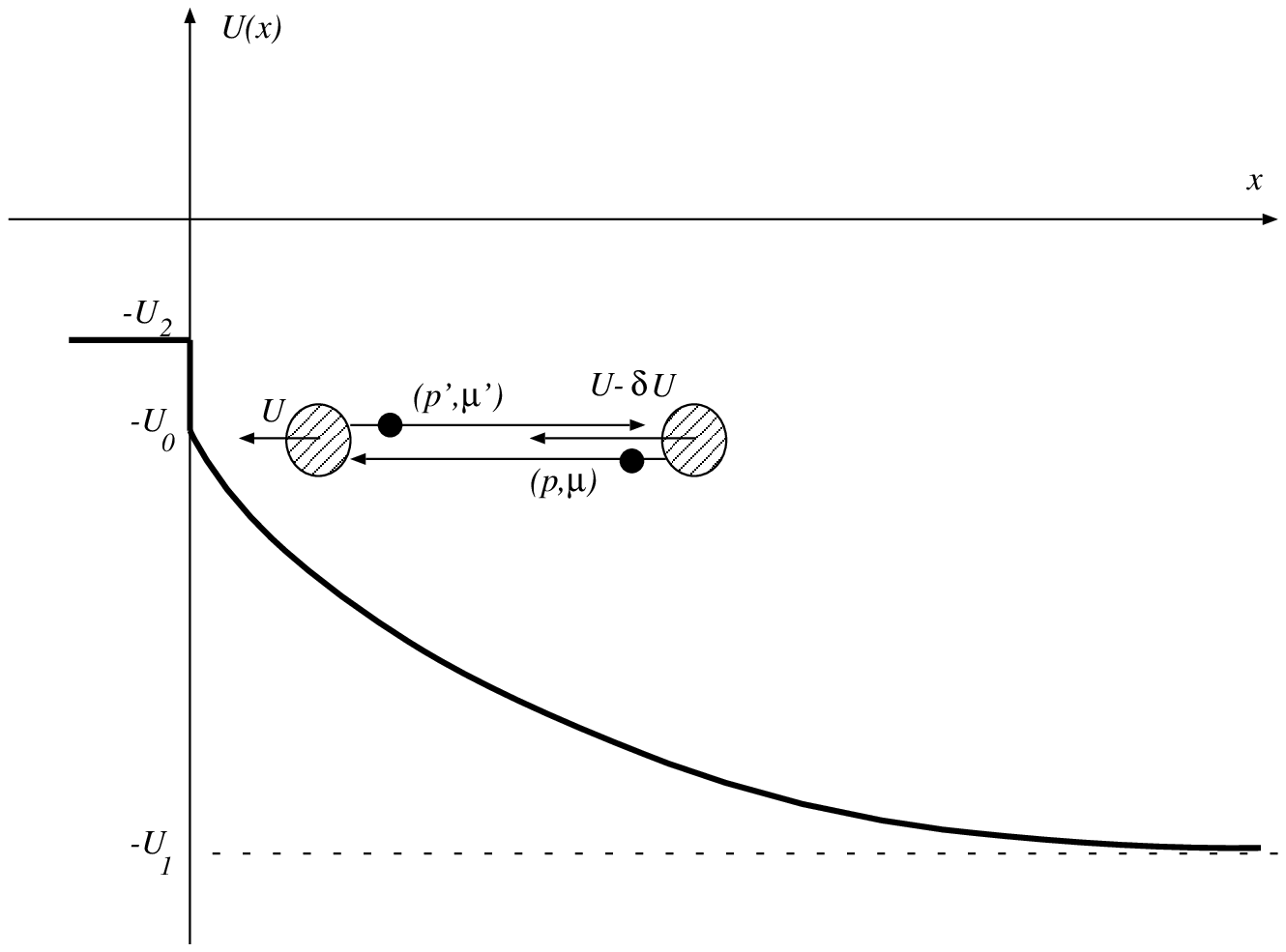}

\caption{Flow velocity profile in a CR dressed shock (solid curve). Hatched
circles represent two scattering centers moving with the flow and
approaching each other at a speed $\delta U.$ The filled circle depict
a CR particle after the first and second collision (see text).\label{cap:Flow-velocity-profile}}
\end{figure}

The situation changes fundamentally, when the number and energy of
accelerated particles increase. First of all, the shock gets dressed
with a cloud of accelerated particles (CRs) diffusing ahead of it,
and the plasma flow becomes significantly modified by their backreaction.
In addition to an abrupt velocity jump (which can be significantly
reduced by this backreaction) the flow develops an extended CR precursor
(CRP, thereafter) of the length $L_{p}\sim\kappa\left(p_{*}\right)/U_{1}=L_{dif}\left(p_{*}\right)$,
where $p_{*}$ is the particle momentum corresponding to the maximum
contribution to the pressure of accelerated particles. The flow in
the CRP gradually slows down towards the main shock (subshock) and
the spectrum is flatter than $p^{-4}$ at high momenta, so that $p_{*}\simeq p_{max}$
and thus $L_{p}\simeq L_{dif}\left(p_{*}\right)$. Let us assume that
a particle undergoes subsequent scattering by two scattering centers,
approaching each other at a speed $\delta U\ll c$, Fig.\ref{cap:Flow-velocity-profile}.
Suppose that immediately after scattering off the right center, the
particle has (in that center reference frame) the momentum $p$ and
the cosine of the pitch angle with respect to the plasma flow (moving
to the left), $\mu<0$. After scattering off the left center, the
particle momentum becomes $p^{\prime}$ and the cosine of the pitch
angle $\mu^{\prime}>0$, \ie the particle moves back towards the
first scattering center. Since scattering is elastic, from particle
kinematics we obtain

\begin{equation}
\frac{p^{\prime}}{p}\simeq\left(1-\frac{\delta U}{c}\mu\right)\left(1+\frac{\delta U}{c}\mu^{\prime}\right).\label{eq:pprime}\end{equation}
This result is obviously identical to the momentum gain in the conventional
DSA theory \citep{bell78a}. The left scattering center can be identified
with one of the downstream, and the right one with one of the upstream,
scattering centers. Considering $\mu$ and $\mu^{\prime}$ as independent
stochastic variables evenly distributed over the intervals $\left(-1,0\right)$
and $\left(0,1\right)$, respectively, one can calculate an average
momentum gain

\begin{equation}
\frac{\left\langle \delta p\right\rangle }{p}\simeq2\frac{\delta U}{c}\left\langle \mu\right\rangle =\frac{4}{3}\frac{\delta U}{c}\label{eq:dpav}\end{equation}
where 

\[
\left\langle \cdot\right\rangle =\int_{0}^{1}\left(\cdot\right)\mu d\mu/\int_{0}^{1}\mu d\mu\]
Note, that the last result also holds in the case when there is an
intervening scattering that does not reverse the particle direction.
This can be seen from eq.(\ref{eq:pprime}) by noting that in such
a case, both $\mu$-s would be of the same sign and cancel upon averaging
to the first order in $\delta U/c$. Just as in the conventional (linear)
acceleration theory a pair of scattering centers does not need (and
is unlikely) to be the same at the next acceleration cycle. However,
this is quite possible in the scattering environment considered later
in this paper. If so, then the momentum gain per cycle $\delta p$
can be obtained from the conservation of the adiabatic invariant

\[
J_{\parallel}=\oint p_{\parallel}dl_{\parallel}=\const\]
where the integral runs between the two scattering centers. If the
interaction time of a particle with a scattering center is much shorter
than the flight time between collisions, then $J_{\parallel}\simeq p_{\parallel}l=\const$,
where $l\left(t\right)$ is the (decreasing) distance between the
two centers. The formula (\ref{eq:dpav})  immediately follows from
the last relation.

\begin{figure}
\plotone{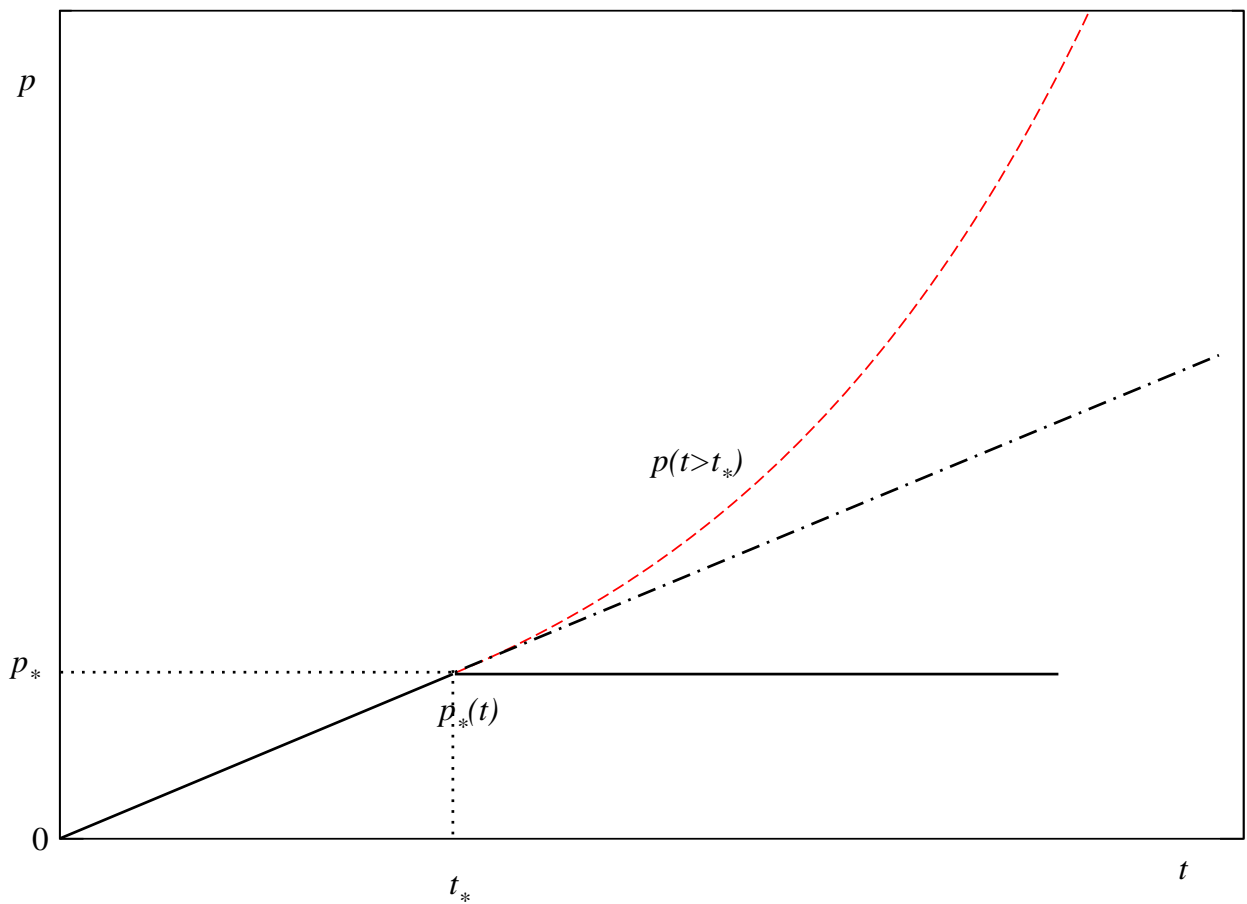}

\caption{Schematic representation of acceleration process. Up to the time
$t_{*}$ the acceleration proceeds at the standard rate, which is
similar during both the linear and nonlinear phases (solid line).
The particle pressure dominant momentum $p_{*}\left(t\right)$
grows as the maximum momentum of the entire spectrum. If for $t>t_{*}$
the pressure dominant momentum $p_{*}$ remains flat (solid line),
particles with $p>p_{*}$ maintain the same acceleration rate that they had
when their momentum was equal to $p_{*}$ (dashed line). The dash-dotted
line shows for comparison, how the maximum momentum $p_{max}=p_{*}$
would continue to grow after $t=t_{*}$, had $p_{*}\left(t\right)$
not stopped growing at $t=t_{*}$.\label{cap:intro}}
\end{figure}

The average time needed to complete the cycle is

\[
\left\langle \delta t\right\rangle =2\left\langle \frac{\lambda}{c\mu}\right\rangle =4\frac{\lambda}{c}\]
where we substituted $\lambda$ for an averaged distance between scattering
centers. Now we can write

\[
\frac{dp}{dt}\equiv\dot{p}\simeq\frac{\left\langle \delta p\right\rangle }{\left\langle \delta t\right\rangle }=\frac{1}{3}p\frac{\delta U}{\lambda}\]
As long as $\lambda\ll L_{p}$, one can approximate $\delta U$ as

\[
\delta U\simeq-\lambda\frac{\partial U}{\partial z}\]
so that the local acceleration rate is

\begin{equation}
\dot{p}\simeq-\frac{1}{3}p\frac{\partial U}{\partial z}\label{eq:pdot}\end{equation}
 We observe that the particle m.f.p. dropped out of the acceleration
rate in the smooth part of the shock structure, in contrast to the
linear acceleration regime which occurrs at the shock discontinuity.
The last formula is, of course, consistent with the standard diffusion
convection equation

\begin{equation}
\frac{\partial f}{\partial t}+U\frac{\partial f}{\partial z}-\frac{\partial}{\partial z}\kappa\frac{\partial f}{\partial z}=\frac{1}{3}\frac{\partial U}{\partial z}p\frac{\partial f}{\partial p}\label{eq:difconv}\end{equation}
if one considers its characteristics ignoring diffusion effects.%
\footnote{As it was shown by \cite{mdru01}, there is a reason to do so in certain
situations. In particular under a strong nonlinear shock modification
and Bohm diffusivity, ($\kappa\propto p$), the velocity profile is
linear, \ie $\partial U/\partial z\simeq const$ in a significant
part of the CRP.%
} Then, in the vicinity of the (abrupt) momentum cut-off $p_{max}$,
the main terms in eq.(\ref{eq:difconv}) are the first one on the
left hand side and the acceleration term on the right hand side. This
clearly shows that a (front) solution propagates in momentum space
at the speed given by eq.(\ref{eq:pdot}), basically independent of
$z$ and $\kappa$. It is important to emphasize, however, that the
independence of the acceleration rate of the particle diffusion coefficient
(or m.f.p., as shown above) does not mean that the acceleration is
faster than in the linear case where such dependence is the main factor
making the acceleration relatively slow. The reason is very simple.
The derivative $\partial U/\partial z$ in eqs.(\ref{eq:pdot}) and
(\ref{eq:difconv}) itself depends on $\kappa$. Indeed estimating
$\partial U/\partial z$ as 

\begin{equation}
\frac{\partial U}{\partial z}\simeq\frac{U_{1}}{L_{p}}\simeq\frac{U_{1}^{2}}{\kappa\left(p_{*}\right)}\label{eq:nlrate}\end{equation}
one sees that the acceleration time is $\tau_{acc}\sim\kappa\left(p_{*}\right)/U_{1}^{2}$
\ie particles with momenta $p\sim\textrm{p}_{*}$ are accelerated
approximately at the same rate as in the linear theory  \citep[apart from
a numerical factor $\mathcal{O}(1)$, see ][]{mdru01}. Clearly,
since the local (\ie inside of the CRP) acceleration rate is independent
of momentum, particles with $p<p_{*}$ accelerate more slowly than
the linear theory acceleration rate, while particles with $p>p_{*}$
accelerate faster rate than in linear theory, and may in principle
be accelerated faster than the Bohm rate. The overall acceleration
process is illustrated in Fig.\ref{cap:intro}, depending on whether
or not the pressure dominant momentum $p_{*}$ is stopped from growing
after some $t=t_{*}$. If it is, particle momentum grows exponentially,
otherwise it continues to grow linearly in time. 

There are two problems with realization of this possibility. First,
because in the nonlinear acceleration regime the spectrum at the upper
cut-off is flatter than $p^{-4}$ \citep[it actually flattens to $p^{-3.25}$
at $p_{max}$, ][ in the limit of high Mach number and $p_{max}\gg mc$]{m97a},
$p_{*}$ is clearly identifiable with $p_{max}$. Therefore, unless
the spectrum slope changes at $p\sim p_{*}$, it is difficult to accelerate
particles beyond $p_{*}$. This is why the acceleration rates in both
linear and nonlinear regimes are similarly slow, although for rather
different reasons. In the linear regime, the acceleration rate slows
down with time since the acceleration cycle duration, $\tau_{cycl}$
grows with momentum (as gyro-period does), so the particle momentum
grows only linearly with time. In the nonlinear regime, the width
of the shock grows linearly with $p_{*}$, so that the acceleration
rate (\ref{eq:nlrate}) decreases with $p_{*}$ (not with $p$!) and
$p_{*}$ growth also linearly with time. Clearly, the first requirement
would be to stop $p_{*}$ from growing at some point or at least to
slow down its growth considerably. If $p_{*}=const$, then particles
with $p>p_{*}$ increase their momentum exponentially, eqs.(\ref{eq:pdot},\ref{eq:nlrate}). 

One simple reason for the saturation of $p_{*}\left(t\right)$ is
a geometrical one. The thickness of the CR cloud ahead of the subshock,
$L_{p}\left(p_{*}\right)$ cannot exceed the accelerator size, \ie
some fraction of the shock radius $R_{s}$, in a SNR, for example,
\citep[\eg][]{dru83,ber96}. Thus we may identify $p_{*}$ with the maximum
momentum achievable in SNR but limited by the geometrical constraints
not by the lifetime of SNRs. What is also required here is that, along
with or, independent of this geometrical limit to $p_{*}$, the character
of \emph{particle confinement} to the shock \emph{changes} at $p_{*}$
when $p_{*}$ reaches some critical value. This should lead to a break
in the spectrum at $p\sim p_{*}$ such that its slope at $p>p_{*}$
is steeper than $p^{-4}$ and flatter at $p<p_{*}$. Then the main
contribution to the particle pressure would come from particles \emph{around}
$p_{*}$. 

For example, the spectral break may be initiated by a strong shock
modification by accelerated particles \citep{mdj02}. Namely, Alfven
wave compression in the CR precursor shifts the waves to the short
wavelength end of the spectrum, leaving particles with $p>p_{*}$without
the resonant waves. In fact, some of those particles still remain
resonant with the waves with $k\ge r_{g}^{-1}(p_{*})\equiv k_{*}$
as long as their pitch angle satisfy the condition $p\mu<m\omega_{c}/k_{*}$,
where $\mu$ is the cosine of the pitch angle, $\omega_{c}$ is the
gyro-frequency. At the same time, the number of these resonant particles
rapidly drops with momentum since their pitch angle distribution is
limited by $\left|\mu\right|<m\omega_{c}/k_{*}p$, and they can hardly
form a good pool for the acceleration. Rather, the momentum spectrum
can be shown decay exponentially at $p>p_{*}$.

However, the latter example is based on a weak turbulence picture
where a sharp resonance of particles with waves of random phases determines
particle dynamics. As we demonstrate in the sequel, strong magnetic
field perturbations result in a quite different character of particle
confinement. In the next section we consider a scattering environment
which (though still simplified) retains some important characteristics
of a more realistic strongly unstable turbulent CRP. As discussed
earlier in this section, this turbulence should satisfy the following
two conditions: i.) particles with momenta $p>p_{*}$ should not suffer
catastrophic losses, ii.) their spectrum must nevertheless be steeper
than $p^{-4}$. As we demonstrate in the next two sections, a gas
of relatively weak shocks, that can emerge in a number of ways, provides
the required scattering.

\section{Particle dynamics\label{sec:Particle-dynamics}}

The conventional paradigm of particle transport in shock environment
is the diffusive pitch angle scattering on self-generated, random
phase Alfven waves. The resulting spatial transport is also diffusive
with the coefficient that scales with the wave amplitude as $\kappa\left(p\right)\propto\delta B_{k}^{-2}$,
where the fluctuation wave number $k$ must be set in the resonance
relation with the particle momentum, $kr_{g}\left(p\right)\simeq1$.
While this is the most consistent plasma physics approach that can
be derived from the first principles using the quasi-linear theory
provided that $\delta B\ll B_{0}$, there is no guarantee that it
holds when the waves go nonlinear \citep[\eg][]{vlk84,ab86}. The latter
is the rule rather than an exception in shock environments, which
has been documented via observations, theory and numerical studies.
So, the Earth's bow-shock as well as interplanetary and cometary shocks
reveal a variety of coherent nonlinear magnetic structures. Usually,
they originate upstream as a result of nonlinearly developed instability
of the distribution of ions reflected from the shock in the case of
oblique propagation. If the shock is quasi-parallel the unstable ion
distribution forms by thermal leakage from the downstream plasma.
In cometary plasma, the pick-up ions drive the instability. Very often
the magnetic structures are observed as an ensemble of discontinuities
referred to as shock-trains or shocklets \citep{tsur}. There is no
shortage of theoretical models describing these features, and we briefly
return to them below. Standing somewhat outside of these models, but
perhaps more specific to the nonlinear CR shocks is the acoustic instability
driven by the gradient of the CR pressure in the CRP. It was first
studied by Drury \citep[see \eg][and references
therein]{drufal,zank90,kjr92} and it also evolves into a gas of moderately strong shock
waves propagating towards the main shock (sub-shock) in the CRP. In
what follows in this section, we will assume that the shock precursor
is filled with the gas of such shocks separated by some distance $L$.

Let us consider such a shock train propagating in the CRP. Obviously
the relative speed between the individual shocks and the speed of
the bulk flow is small compared to the particle speed, $U\left(z\right)\ll c$.
Therefore, we can consider the problem of particle scattering by scattering
centers (shocks) and particle acceleration caused by the relative
motion of the scattering centers, separately. Note that the acceleration
part of the problem has been preliminarily considered in sec.\ref{sec:Acceleration-Mechanism}
and we will return to it in sec.\ref{sec:Particle-spectrum-between}.
Turning here to the scattering part of the problem, we assume for
simplicity that the shocks are one-dimensional and propagate at the
same speed along the shock normal. Furthermore, we transform to their
reference frame, neglecting compression by the main flow $U\left(z\right)$
and their relative motion. Even under these simplifications, the shock
train magnetic structures can be still rather complicated. For example,
they have been extensively studied in the frame of the so-called derivative
nonlinear Schr\"odinger equation and its modifications \citep{medvd95}.
Alfven -- ion acoustic wave coupling results in wave steepening and
gives rise to a shocktrain. A kinetic effect included in this model
is particle trapping which was shown to be important for the formation
of the shocktrain, which has been studied numerically \citep{hada,medv97}.
In another model, the shocktrain forms from a balance of a quasi-periodic
driver and the nonlinearity of the magnetic field perturbations. The
driver may originate from one or a few unstable harmonics providing
the smooth parts of the solution that steepen into shocks. These shocks
also have a complicated spatial structure, typically characterized
by a fast rotation of the magnetic field vector, caused by dispersion
\citep[see][and references therein]{mphysD96}. However, since the
gyroradius of high energy particles, even if reduced by possible magnetic
field amplification, is still much larger than both the width of the
shock transition and the dispersive scale $c/\omega_{pi}$, we replace
each shock by a discontinuity of a coplanar magnetic field. A different
possibility that results in a similar behaviour of the magnetic field
occurs when the strong Drury instability discussed earlier, evolves
into an ensemble of moderately strong shocks. 

\begin{figure}
\plotone{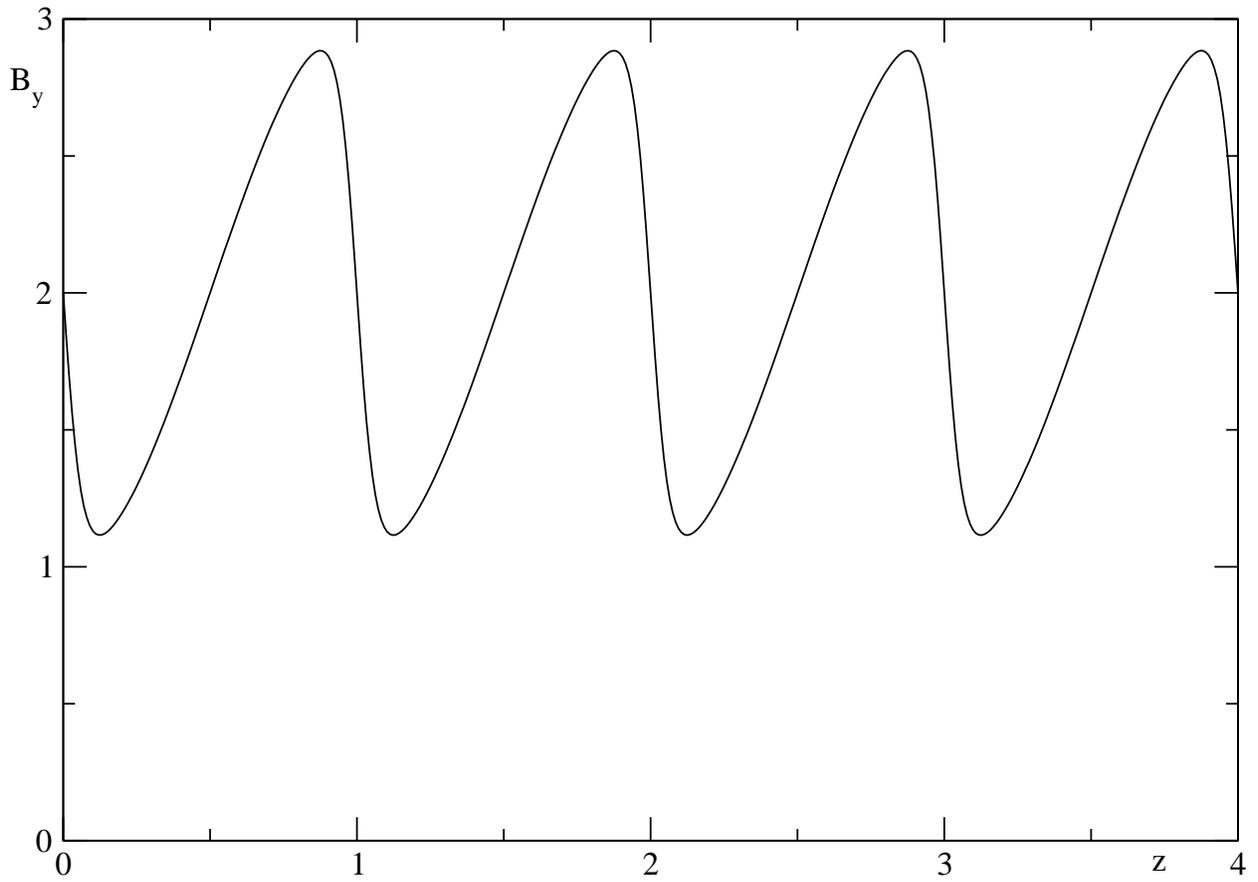}

\caption{$y$-component of magnetic field in the shocktrain given by eq.(\ref{eq:mf})
for $\nu=0.2$, $\tilde{B}_{y}=1$ and $\overline{B}_{y}=2$.\label{cap:mf}}
\end{figure}

Both cases are covered by the following representation of the magnetic
field which corresponds to a periodic sequence of shocks, in which
only one component of the field varies with the coordinate in a co-moving
reference frame, $\mathbf{B}=\left(0,B_{y},1\right)$, where

\begin{equation}
B_{y}\left(z\right)=\overline{B}_{y}+\tilde{B}_{y}\sin\left[\pi\left(z-1/2\right)\right]\tanh\left\{ \frac{1}{\nu}\cos\left[\pi\left(z-1/2\right)\right]\right\} \label{eq:mf}\end{equation}
Here $\overline{B}_{y}$ is the constant component of the transverse
to the shock normal component of the magnetic field and $\tilde{B}_{y}$
characterizes the strength of the shocks in the shocktrain (both normalized
to the constant $z$-component, $B_{0}$), Fig.\ref{cap:mf}. We normalized
the coordinate $z$ to the distance between shocks, $L$. Of course,
in reality different shocks in the shocktrain have not the same strength
so that the coefficient $\tilde{B}_{y}$ should be replaced by a stochastic
variable with a pdf, inferred from the shock dynamics. We return to
this discussion in the next section, but in this simplified study
we assume all the shocks in the shock train have the same strength.

Let us consider particle dynamics in the magnetic field given by eq.(\ref{eq:mf}).
It is convenient to write the equations of motion using dimensionless
variables in which time is normalized to $L/c$. Since we will be
primarily concerned with the dynamics of particles having momenta
$p\ga p_{*}\gg mc$, we normalize their speed to the speed of light
and set the absolute value of particle velocity to unity, $V\approx1$.
The remaining variable $p$ is normalized naturally so as to the corresponding
gyroradius $r_{g}\left(p\right)$ is measured in the units of $L$,
that is the particle momentum $p$ is scaled to $eB_{0}L/c$. We also
introduce the cosine of the pitch angle $\mu=p_{z}/p$. The equations
of motion read 

\begin{eqnarray}
\dot{\mathbf{p}} & = & \mathbf{n}\times\mathbf{B}\label{eq:mot1}\\
\dot{z} & = & \mu\label{eq:mot2}\end{eqnarray}
 where $\mathbf{n}=\mathbf{p}/p$ and $p=const$.

\begin{figure}
\plotone{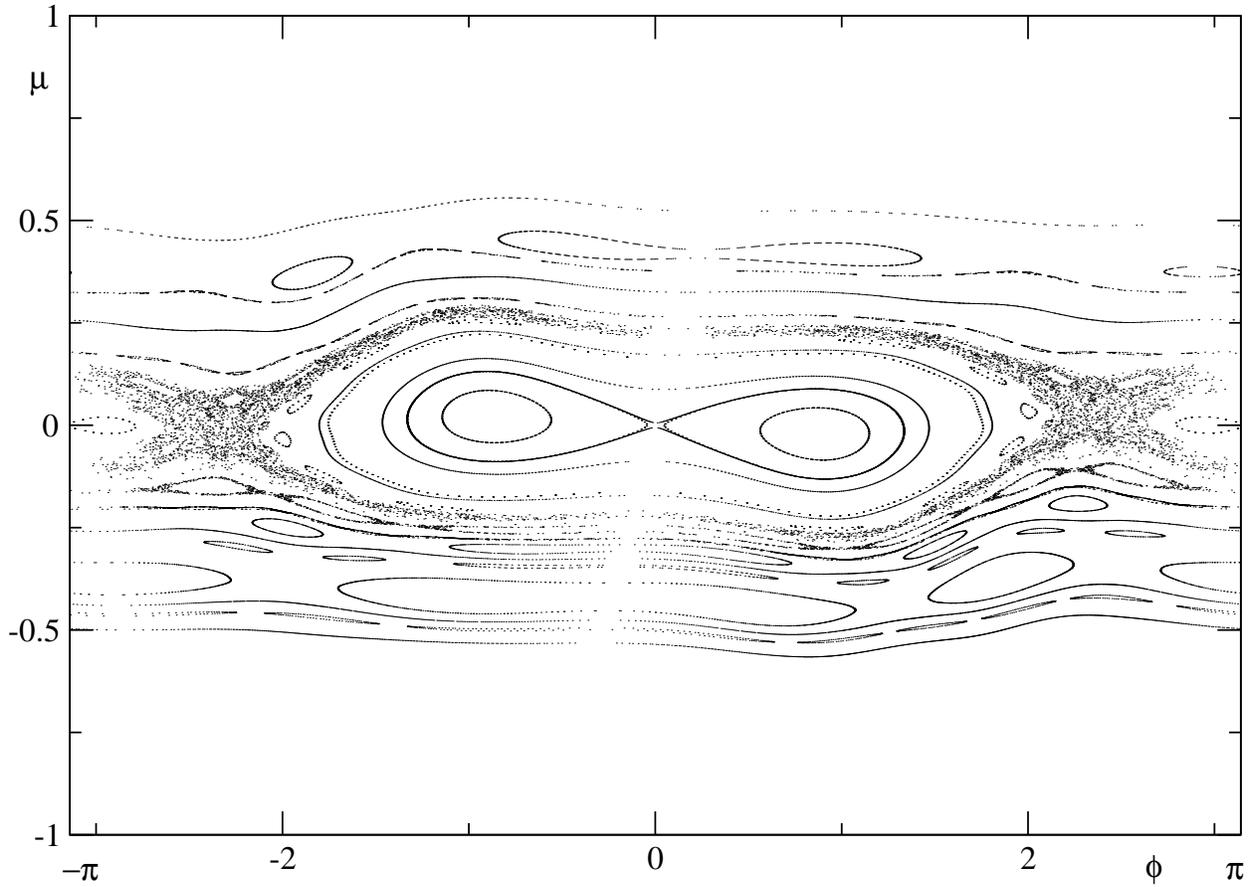}

\caption{Particle orbits shown for $\nu=0.5;\;\overline{B}_{y}=2;\;\tilde{B}_{y}=0.3$
in eq.(\ref{eq:mf}) and $p=1$ on the $\phi,\mu$ plane. Each dot
corresponds to the intersection of particle trajectories with one
of the planes located at an integer $z$ and coinciding with the shock
locations (Poincare section). Several particle orbits form invariant
curves (KAM-tori) as well as stochastic layers around some of the
separatrices (see text).\label{cap:qreg}}
\end{figure}

Fig. \ref{cap:qreg} shows particle trajectories as a Poincare section,
that represents orbit crossings of the shocks, \ie the planes $z=j$
(where $j$ is an integer) on the gyrophase -pitch angle plane, ($\phi,\mu$).
To make connection with the drift approximation we have transformed
the particle momentum to the coordinate system rotated around the
$x$ axis by the angle $\vartheta=\tan^{-1}\overline{B}_{y}$. In this
coordinate system, particles spiral around the $z$ axis with $\mu=\const$
if $\tilde{B}_{y}=0$. The gyroradius is taken to be equal to the
distance between the shocks ($p=1$, resonant case in terms of the
linear theory) but the shock transitions are broad so that no distinct
shocks with the steep magnetic field gradient are present and the
particle dynamics remains largely regular. For smaller $p$ the system
becomes almost integrable since it can be reduced to a two dimensional
one, because the particle magnetic moment is conserved with good accuracy.
In the case shown in Fig. \ref{cap:qreg}, however, some of the separatrices
are clearly destroyed. Nevertheless, there is a clear separation of
trapped (magnetic mirroring) and passing particles. Based on the conservation
of the first adiabatic invariant $I\propto\left(1-\mu^{2}\right)/B$
(magnetic moment), the trapped particles must be confined in momentum
space to $\left|\mu\right|<\mu_{cr}=\sqrt{\left(B_{max}-B_{min}\right)/B_{max}}\approx0.27$
which is close to what can be seen from Fig. \ref{cap:qreg}. It is
important to emphasize here that the trapping area does not shrink
with $p$ as was the case in the linear resonance situation, discussed
at the end of the previous section. Clearly the spatial transport
of these particles is non-diffusive. The trapped particles are convected
with the shocktrain. The untrapped particles ($\left|\mu\right|>\mu_{cr}$)
propagate ballisiticaly and escape from the CR cloud. Thus, even a
small amplitude (coherent) shock train provides a substantially different
confinement of particles than the small amplitude random ensemble
of Alfven waves does. It ensures perfect confinement of trapped particles
and no confinement at all of untrapped particles. This is what is
required for the acceleration mechanism outlined in Sec. \ref{sec:Acceleration-Mechanism}.
When this confinement regime takes over above some $p=p_{*}$, since
there are no resonant waves of the length larger than $r_{g}\left(p_{*}\right)$,
the energy spectrum must become steeper for $p>p_{*}$. We will come
back to this point in Sec.\ref{sec:Particle-spectrum-between}.

The dynamics changes even more dramatically when the thickness of
the shocks in the shocktrain becomes small compared to the particle
gyroradius. Taking the same parameters in eq.(\ref{eq:mf}) as those
used in Fig.\ref{cap:qreg} except for $\nu\approx0$, we show in
Fig.\ref{cap:Chaos} the Poincare section created by a single trajectory.
It should be noted, however, that the actual amplitude of the shocks
also increases somewhat in this case, even though the amplitude parameter
$\tilde{B}_{y}$ is fixed. The phase space is now a {}``stochastic
sea'' with embedded {}``islands'' of quasi-regular motion, quite
typical for the deterministically chaotic systems \citep[see \eg ][]{lib:licht}.
The dynamics inside the islands is weakly stochastic, so that the
particle trajectories there are closer to the nearly integrable case,
shown in Fig.\ref{cap:qreg}, than in the rest of the phase plane.
Therefore, within those islands where the averaged pitch angle $\overline{\mu}\neq0$,
particles propagate ballisticaly in the $z$ direction, similar to
the untrapped particles in the previous example. There are also islands
where $\overline{\mu}\approx0$, and particle propagation is substantially
suppressed there. The islands are stochastic attractors and the mapping
exhibits cycling around them apart from the quasi-periodic motion
inside islands, not shown in Fig. \ref{cap:Chaos}. Note that only
a single trajectory is shown. The first type of islands ($\overline{\mu}\neq0$)
are responsible for Levy flights (ballistic mode of particle propagation)
whereas the second type of islands ($\overline{\mu}=0$) represent long
rests or traps. The rest of the phase plane is covered by the region
of a global stochasticity where particle propagation appear to resemble
that around the islands but with shorter rests and Levy flights. Note
that we use the term Levy flights for the above transport events even
though the probability distribution function of the jump lengths might
have a finite first moment (mean jump length). We investigate particle
propagation in the $z$-direction in more detail in the next section.

\section{Particle transport\label{sec:Particle-transport}}

It is useful at this point to recall that in the standard diffusive
shock acceleration the pitch-angle diffusion determines the spatial
transport which is also diffusive. This picture results from the scattering
of particles by Alfven waves of small amplitude and random phases.
By contrast, in the preceding section we have considered a coherent
magnetic structure of the finite amplitude shocks, as an alternative
particle scattering agent at momenta higher than the pressure containing
momentum scale $p_{*}$. Despite the coherence, the particle dynamics,
as we have seen, remains strongly chaotic and the onset of stochasticity
depends on both the amplitude of the shocks and their thickness. We
thus appeal here to the intrinsic stochasticity, caused by unpredictable
particle motion in a field which can be perfectly regular. The fact
that in a real situation the field is also random is often assumed
to be of secondary importance. 

\begin{figure}
\plotone{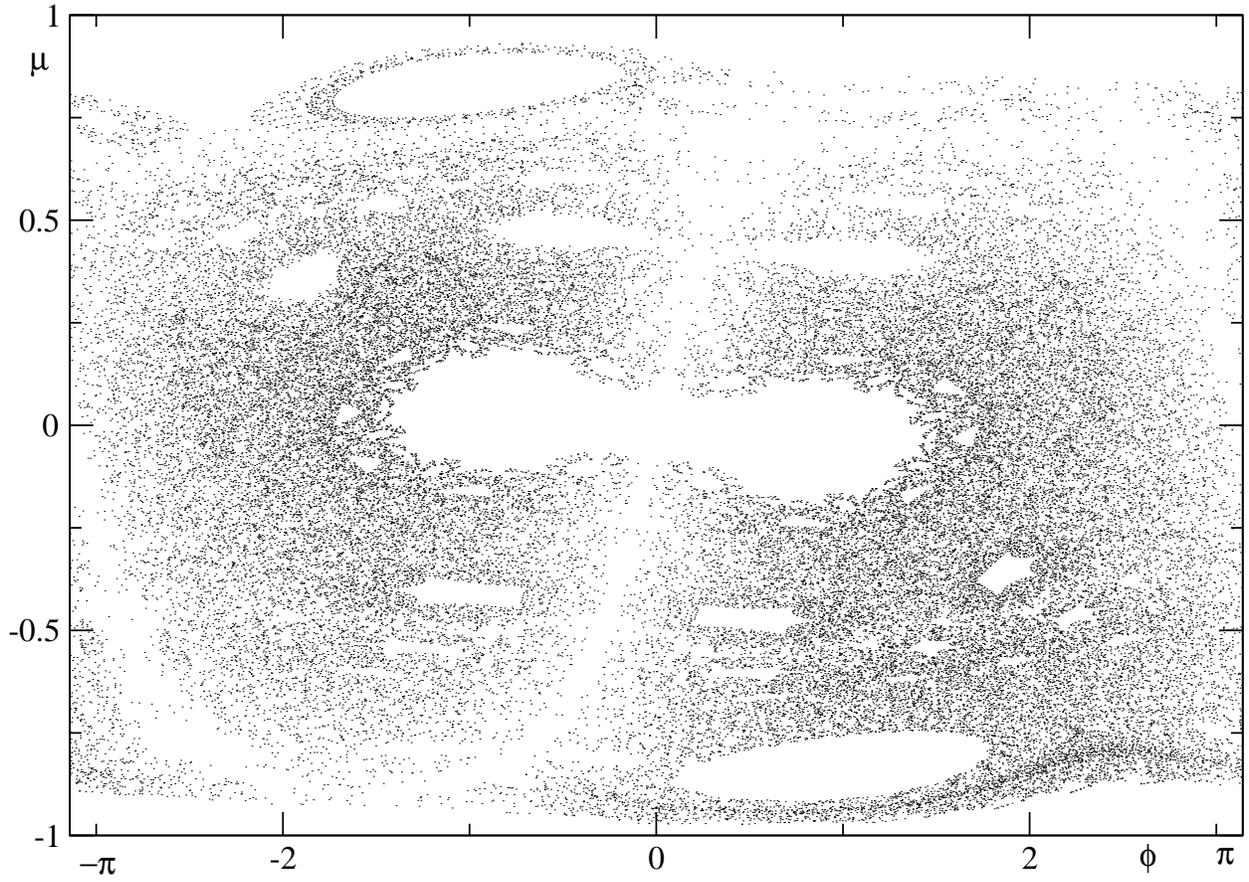}

\caption{The same as Fig. \ref{cap:qreg} except for the narrow shock transitions,
$\nu\approx0$. In contrast to Fig.\ref{cap:qreg}, however, the entire
phase portrait is formed by a single particle orbit, that does not
visit the 'holes' in the phase space.\label{cap:Chaos} }
\end{figure}

The resulting spatial transport is not a {}``classic'' diffusion
process. The connection between the particle dynamics, represented
by the Poincare section shown in Fig.\ref{cap:Chaos} and the particle
spatial transport can be most easily understood by considering eq.(\ref{eq:mot2}),
$\dot{z}=\mu$ as a stochastic differential equation, in which $\mu$
is a random process. Some general characteristics of this process
can be inferred from the Poincare section shown in Fig.\ref{cap:Chaos}.
As we mentioned above there are traps represented by islands which
translate in the long flights or long rests depending on whether the
averaged value of $\mu$, \ie $\overline{\mu}$ is nonzero or zero, respectively.
Here the averaging is to be taken over an island attractor which is
a layer of enhanced phase density around the island. 

The particle transport can be conveniently treated as a random walk
on a lattice of shocks located at the integer values of $z$. However,
in contrast to the classical random walk, this is a non-Markovian
process since it is characterized by long rests \citep{zasl02}. Particles
interact with the same shock repeatedly while they are trapped near
an island with $\overline{\mu}=0$. Similarly, they perform a long jump
in one direction and cross many shocks in a row while they are trapped
in the phase space around an island with $\overline{\mu}\neq0$. 

\begin{figure}
\plotone{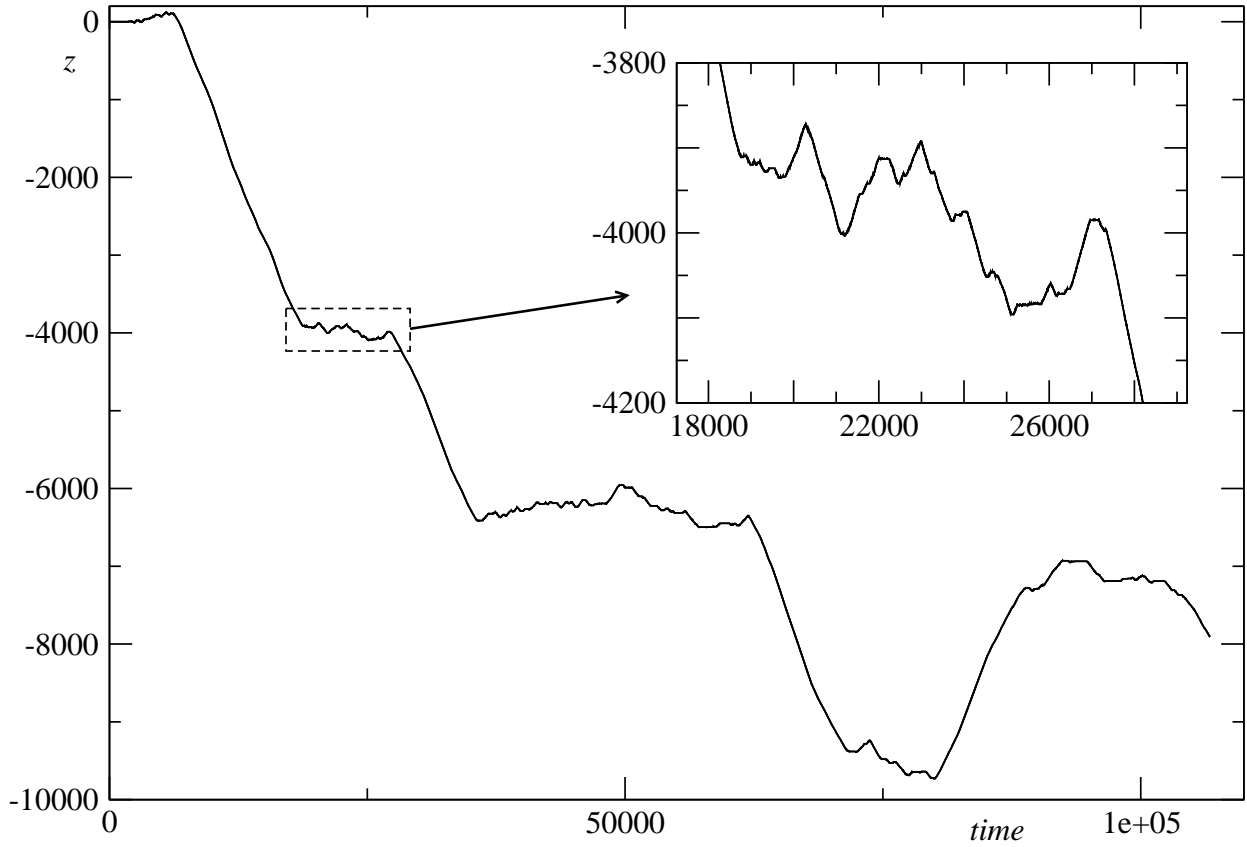}

\caption{Particle trajectory represented as $z\left(t\right)$ that corresponds
to the phase space shown in Fig.\ref{cap:Chaos}. It starts close
to the islands near $\mu=0$ so that the orbit stays close to the
origin ($z=0$) for a long time. Such long rests will be repeated
many times at different locations. Overall, particle transport is
organized in clusters (five such clusters are shown) where the transport
is strongly suppressed. The clusters are connected with long jumps
where the transport is ballistic.\label{cap:zOft} }
\end{figure}

The particle trajectory that corresponds to the Poincare section shown
in Fig.\ref{cap:Chaos}, is demonstrated in Fig.\ref{cap:zOft}. The
trajectory is indeed, non-diffusive. It consists of long 'Levy flights'
that connect areas (clusters) of a rather slow particle propagation.
Such clustered propagation regime is quite typical for nonlocal (fractional)
diffusion models \cite[see \eg ][]{metzler}. The propagation within
a typical cluster is magnified and shown in the inset of Fig.\ref{cap:zOft}.
Certain similarity with the dynamics at large can be noted, which
is also not unusual for the deterministically chaotic dynamical systems.
However, the intra-cluster Levy flights are significantly shorter
than the inter-cluster ones, the effect that will be shown to be much
stronger for stronger shocks. Besides, long rests at different shocks
can be seen inside the clusters. Note that it is this intra-cluster
particle dynamics that ensures their confinement to a localized region
of the flow convected across the CRP as described in Sec.\ref{sec:Acceleration-Mechanism}.
As we argued there, every such passage through the CRP approximately
doubles the particle momentum. The Levy flights, on the other hand,
can result in the escape of a particle from the shock precursor. We
quantitatively describe and compare both possibilities in Sec.\ref{sub:Acceleration-model:-details}.

\begin{figure}
\plotone{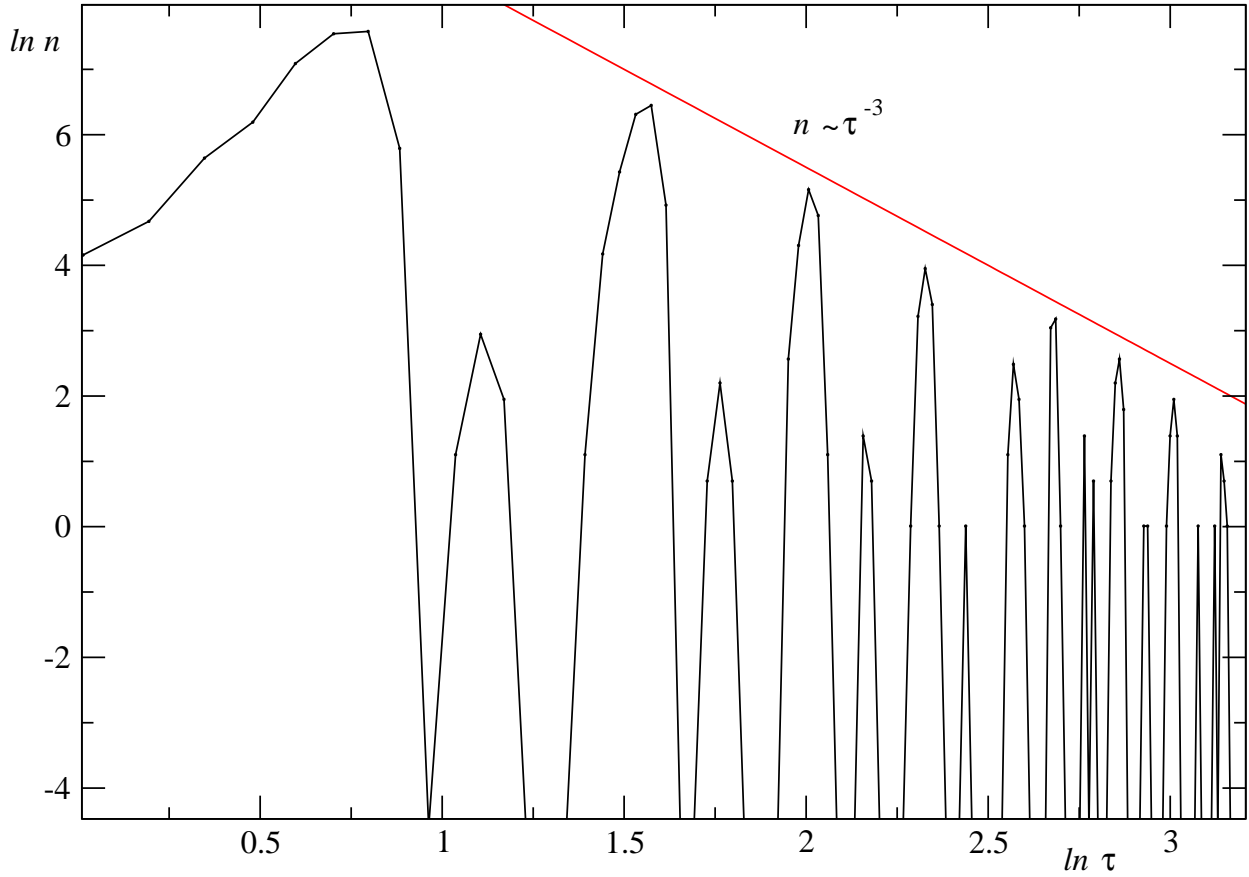}

\caption{Waiting time distribution in a log-log format. Here $n$ is the number
of times the orbit has spent time $\tau$ at any shock in the shocktrain.
A somewhat discrete character of distribution is related to the gyromotion
of particles near a shock so that when they interact with the shock
repeatedly, the interaction time is commensurate with the gyro-period.\label{cap:nOfTau} }
\end{figure}

\begin{figure}
\plotone{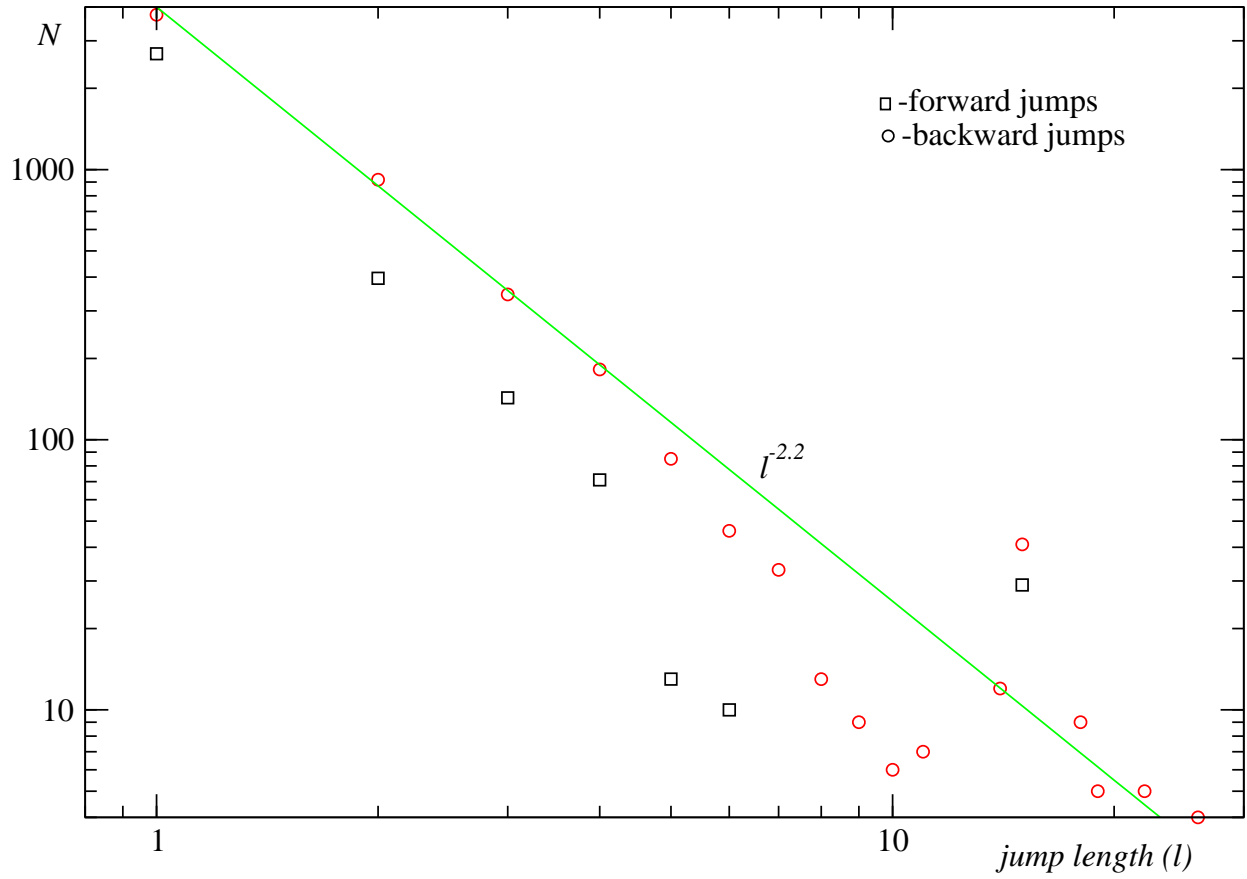}

\caption{Distribution of Levy flights shown for forward and backward jumps.\label{cap:Nofl}}
\end{figure}

\begin{figure}
\plotone{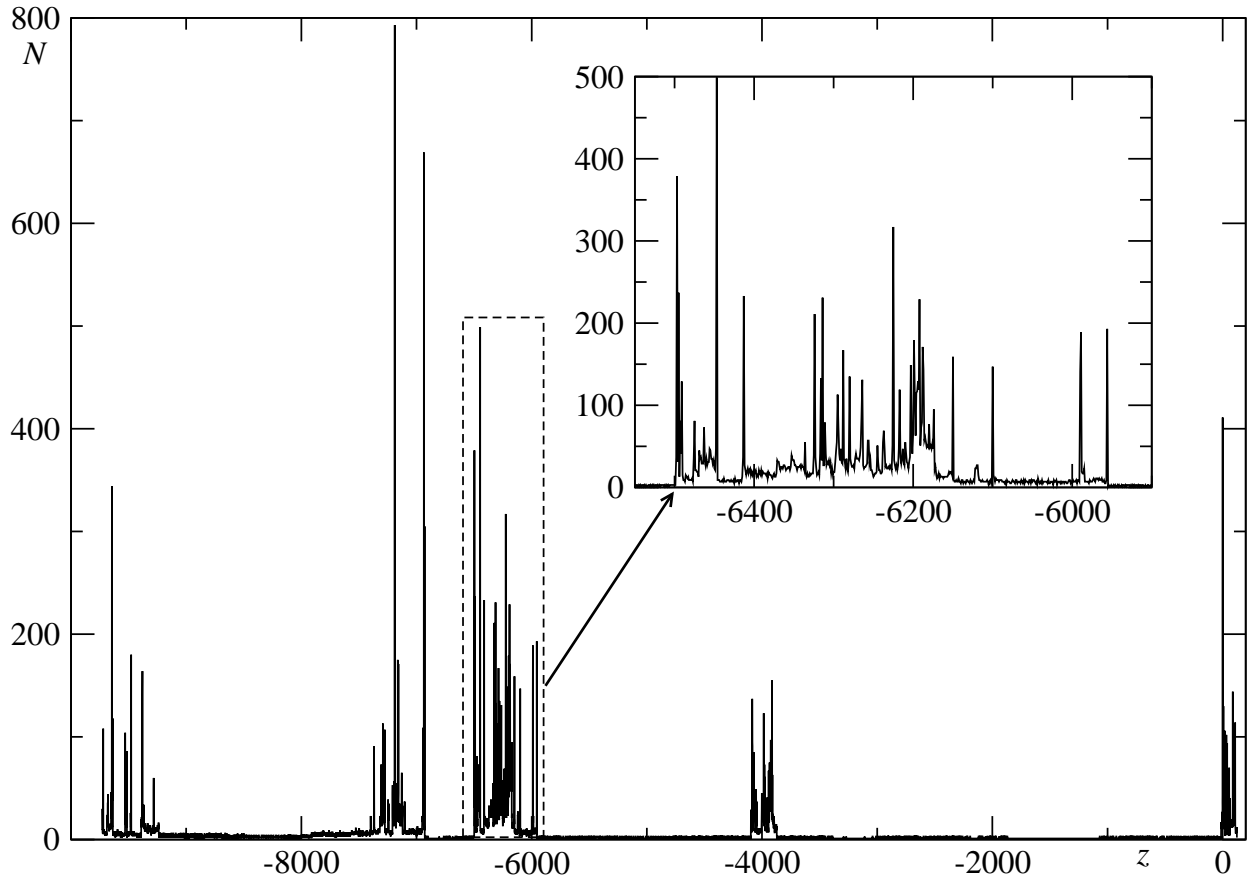}

\caption{Distribution of number of visits of different shocks in the shocktrain.
The magnified box shows the distribution inside of one of the clusters
(see text).\label{cap:NofZ}}
\end{figure}

Let us come back to the overall dynamics. There are two important
characteristics that describe the random walk process. These are the
waiting time and the jump length probability density functions (pdf).
We have obtained these quantities from the trajectory shown in Fig.\ref{cap:zOft}.
The waiting time distribution at different shocks is shown in Fig.\ref{cap:nOfTau}.
Due to the gyro-motion of particles near the shocks, the pdf is not
smooth but it has a smooth envelope which decays approximately as
$\tau^{-3}$, where $\tau$ is the waiting time. The distribution
of the lengths of the Levy flights is presented in Fig.\ref{cap:Nofl}.
Because of the asymmetry of the shock train (Fig.\ref{cap:mf}) there
is a distinct, directional asymmetry of the jumps. Otherwise they
exhibit an approximate power-law distribution up to the jumps having
lengths limited by about $10$. The pdf describes both the intra-cluster
and inter-cluster particle jumps but only the jumps of the first type
are suitable for the continuous acceleration. The longer jumps, as
we estimate in sec.\ref{sec:Estimate-of-the maxmom}, generally result
in the loss of particles. Another way of representing the trajectory
clustering is illustrated in Fig.\ref{cap:NofZ}, where the distribution
of numbers of visits of different sites in the shocktrain is shown. 

The knowledge of the waiting time and jump distributions is necessary
to derive the form of the adequate operators in the transport equation,
 \citep[\eg][]{metzler}. Due to the non-Markovian character of transport
these operators are integral (also called fractional differential
operators, due to the memory in stochastic trajectories) rather than
conventional diffusion operators. In sec.\ref{sec:Particle-spectrum-between}
we shall take a somewhat simpler approach, which, however also requires
the flight- rest time pdf to determine the transport coefficients. 

\begin{figure}
\plotone{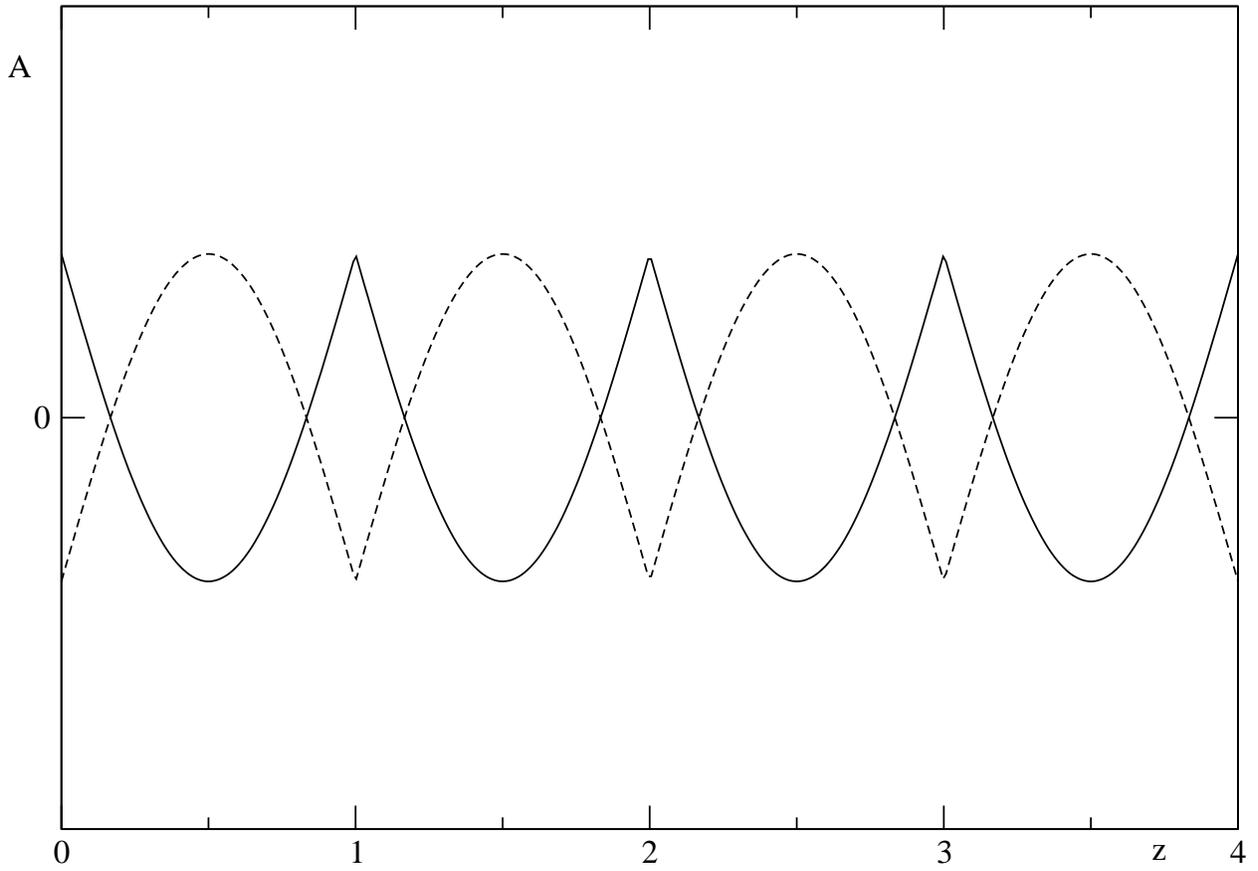}

\caption{Potential $A\left(z\right)\cos\left(\phi\right)$ in eq.(), corresponding
to the magnetic field $B_{y}=-\partial A/\partial z$, given by eq.(\ref{eq:mf})
with $\nu=0$ and $\overline{B}_{y}=0$. The function $A\left(z\right)=\left(\tilde{B}_{y}/\pi\right)\left[\cos\pi\left(z-1/2\right)-1/2\right]$.
Solid line: $\cos\left(\phi\right)=1$, dashed line: $\cos\left(\phi\right)=-1$.
\label{cap:OscLat}}
\end{figure}

Up to now we have numerically examined the particle transport in shocktrains
of a period longer or equal to the particles gyroradius. This corresponds
to the resonant wave-particle interaction in the small amplitude limit.
Obviously even in the case $p\ll1$, there are resonant Fourier components
in the shocktrain, since its spectrum decays as $k^{-2}$ for $k>2\pi$
(recall that the shocktrain period is set to unity). This is illustrated
by Fig.\ref{cap:zOft2}, where particle trajectories with smaller
momenta are shown. It should be also noted, however, that particles
with smaller gyro-radii, such as $p\la p_{*}$ can be efficiently
confined due to the resonance with self-generated waves since, as
we argued earlier, their population is significantly more abundant
than those with $p>p_{*}$ because of the steeper spectrum in this
momentum range. This result will be confirmed later. It is clear that,
no matter how long the period of the shocktrain is, we need to estimate
the confinement of particles with a gyroradius larger than that period,
\ie those with $p>1$ . One can expect that their confinement will
be progressively deteriorated due to the fact that there will be only
a high frequency force acting on these particles. To estimate the
fraction of $p\gg1$ particles that can be confined, we write eq.(\ref{eq:mot1})
in the form of the following system

\begin{eqnarray}
\dot{\phi} & = & -\frac{1}{p}+\frac{\tilde{B}_{y}\left(z\right)\mu}{p\sqrt{1-\mu^{2}}}\sin\left(\phi\right)\label{eq:fidot}\\
\dot{\mu} & = & \frac{\tilde{B}_{y}\left(z\right)}{p}\sqrt{1-\mu^{2}}\cos\left(\phi\right)\label{eq:mudot}\\
p & = & \const\nonumber \end{eqnarray}
where we have assumed for simplicity that $\overline{B}_{y}=0$. Because
$\tilde{B}_{y}\left(z\right)$ has zero average, $\phi$ and $\mu$
are particle gyro-phase and the pitch angle that are taken with respect
to the shock normal. Assuming $\mu\ll1$ (which we verify below),
from eqs.(\ref{eq:mot2},\ref{eq:mudot}) we then obtain

\begin{equation}
\ddot{z}-\frac{1}{p}\tilde{B}_{y}\left(z\right)\cos\left(\phi\right)=0.\label{eq:zdd}\end{equation}
Considering $\phi$ as slowly varying compared to $z$ and $\mu$,
we can integrate the last equation once

\begin{equation}
\frac{p}{2}\dot{z}^{2}+A\left(z\right)\cos\left(\phi\right)=\const+O\left(p^{-1}\right)\label{eq:enint}\end{equation}
where we have introduced a periodic function $A\left(z\right)$ according
to $\tilde{B}_{y}\left(z\right)=-\partial A/\partial z$. Since $\phi$
is a slowly varying variable (it varies on a time scale $p\gg1$,
while $z$ and $\mu$ vary on a time scale $\sqrt{p}$), eq.(\ref{eq:enint})
describes an oscillator lattice with a slowly varying potential as
shown in Fig.\ref{cap:OscLat}. The factor $\cos\left(\phi\right)$
in eq.(\ref{eq:enint}) slowly inverts the lattice potential. However,
trapped particles do not detrap completely but rather get trapped
in one of the nearby potential wells. The width of the trapping zone
in $\dot{z}$ and thus in $\mu$ is given by $\left|\mu\right|<\mu_{s}$,
where 

\[
\mu_{s}=\sqrt{\frac{2}{p}\left(A_{max}-A_{min}\right)}\ll1\]
For $\tilde{B}_{y}\left(z\right)$ given by eq.(\ref{eq:mf}), we
obtain: $A_{min}=-A_{max}$ and $A_{max}=\tilde{B}_{y}/2\pi$. Particles
with $\left|\mu\right|>\mu_{s}$ are not confined to the shocktrain
and propagate ballisticaly.

\begin{figure}
\plotone{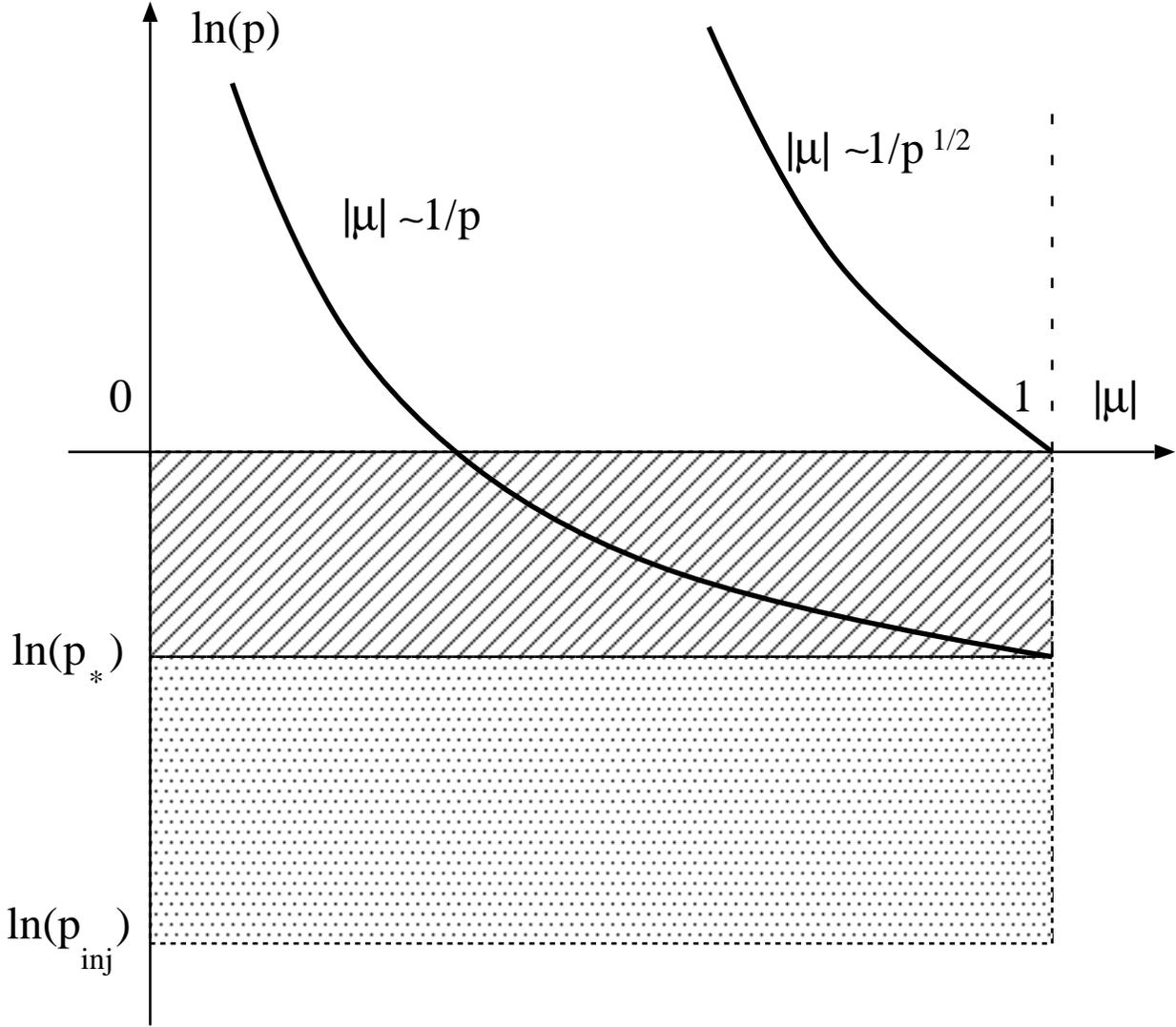}

\caption{Phase space of particles in pitch-angle-momentum representation.
The shaded area ($p_{inj}<p<p_{*},\;\left|\mu\right|<1$) corresponds
to the conventional particle confinement via randomly phased Alfven
waves. For higher momenta, $p>p_{*}$, {[}if there were only weakly
turbulent waves present with minimum wave number, $k=k_{min}=1/r_{g}\left(p_{*}\right)${]}
this type of particle confinement would be limited to $\left|\mu\right|<\mu_{c}\propto1/p$,
because of the resonance condition $kr_{g}\left(p\right)\mu=1$. A
shock-train with the spacing $r_{g}\left(p_{*}\right)<L<r_{g}\left(1\right)$
confines particles similar to the mirroring-type confinnement $\left|\mu\right|<\mu_{c}=\sqrt{\Delta B/B}$
(hatched area), except for the phase space fragmentation (see Fig.\ref{cap:Chaos}
and text). Beyond $p=1$ the particle confinement deteriorates to
$\left|\mu\right|<\mu_{s}\propto1/\sqrt{p}$ (see text).\label{cap:Phase-space-of}}
\end{figure}

To conclude this section, we emphasize that when the momentum of particles
grows and their Larmor radius increases beyond the distance between
the shocks, the fraction of particles that \emph{do not} escape ballisticaly
shrinks with momentum as $\mu_{s}\propto1/\sqrt{p}$, Fig.\ref{cap:Phase-space-of}.
This is slower than in the standard, weakly turbulent picture, where
the critical value of $\mu$ decays as $1/p$. On the other hand,
the latter dependence is based on the linear cyclotron resonance without
broadening, while the resonance broadening also improves the confinement
\citep{acht81}. In any case, one has to expect the spectral cut-off
at $p\sim1$ in units used in this section (Larmor radius is of the
order of shock spacing). Therefore, the characteristic distance between
shocks determines the maximum momentum of accelerated particles. Of
course, in a real situation there is a continuous strength/distance
distribution of shocks which should produce a smoother spectrum decay
at the cut-off momentum. Here we merely estimate the maximum distance
which is equivalent to the minimum wave number in the standard acceleration
picture. In the framework of the acceleration mechanism considered
in this paper, the minimum wave number of the randomly phased Alfven
waves should be of the order of $r_{g}^{-1}\left(p_{*}\right)$, while
particles with the momenta $p_{*}<p<1$ are confined via the mechanism
of interaction with the shocktrain. In the next section we estimate
the distance $L$ between the shocks and thus $p_{max}$ since $r_{g}\left(p_{max}\right)\simeq L$,
again, similarly to the standard estimate $r_{g}\left(p_{max}\right)\simeq k_{min}^{-1}$.

\section{Estimate of the maximum momentum\label{sec:Estimate-of-the maxmom}}

In this section we estimate the maximum distance between the shocks
in the shocktrain which, according to the preceding section, is directly
related to the maximum momentum. In what follows we assume that the
shocks are formed due to the development of an acoustic instability
driven by the CR pressure gradient in the CRP. The latter has the
scale height $L_{p}$ as being created by particles accelerated in
a standard DSA manner and having maximum momentum $p_{*}$. Specifically,
$L_{p}\simeq\kappa\left(p_{*}\right)/U_{1}$. As it was demonstrated
by \cite{drufal}, the linear growth rate of this instability can
be written as

\begin{equation}
\gamma_{D}^{\pm}=-\frac{\gamma_{C}P_{C}}{\rho\kappa}\pm\frac{P_{Cz}}{C_{s}\rho}\left(1+\frac{\partial\ln\kappa}{\partial\ln\rho}\right)\label{DrGrRate}\end{equation}
 Here $P_{C}$ and $P_{Cz}$ are the CR pressure and its gradient
(which is antiparallel to  the shock normal), respectively, $\gamma_{C}$
is their adiabatic index, $\rho$ is the plasma mass density and $C_{s}$
is the sound speed. For efficiently accelerating shocks, one can assume
$\gamma_{C}\approx4/3$.  The first term in eq.(\ref{DrGrRate}) represents
the wave damping caused by CR diffusion, calculated earlier by \cite{ptuskin81}.
The second term is positive for the waves propagating along the pressure
gradient while the oppositely propagating waves are damped. The factor
in the parentheses can reduce or even completely eliminate the instability,
if $\partial\ln\kappa/\partial\rho=-1$. However, there is no physical
grounds, upon which this particular selection should be made \citep{drufal}.
It should be also noted here that particle diffusivity $\kappa$ refers
here to particles with $p\la p_{*}$ while the transport of higher
energy particles is nondiffusive.

Within the approximation leading to the growth rate given by eq.(\ref{DrGrRate}),
there is no dependence on the wave number. A more thorough investigation
can be found in \cite{kjr92}, which shows that the wave number dependence
is indeed not strong. Under these circumstances it is reasonable to
assume that the largest seed waves are the most important ones. Further
we assume that the latter are related to the cyclotron instability
of the Alfven and magnetosonic waves (which can provide the compressional
seed component) that have excited by the accelerated particles with
momenta $p\la p_{*}$, since particles of higher momenta have a steeper
spectrum and do not contribute significantly to the growth rate of
the cyclotron instability. We should focus primarily on the furthermost
part of the precursor, since perturbations starting to grow there
have the best chances to develop fully while convected with the flow
towards the subshock. This part of the precursor is accessible only
to the particles near the maximum momentum achievable within the standard
acceleration scenario, \ie  $p_{*}$. Hence, we assume that the typical
wavelength of the Drury instability is $\lambda_{D}\sim r_{g}\left(p_{*}\right)$. 

While these perturbations grow to a nonlinear level, and propagate
with the flow towards the subshock, they also steepen into the shocks.
This behavior is seen in both two fluid simulations of \cite{drufal}
and, what is particularly relevant to our study, in the kinetic simulations
of \cite{kjr92}. Due to the significant difficulties in numerical
realization of the kinetic model, this study is limited in the maximum
particle momentum, and thus in the length of the CRP, $L_{p}$. Besides
that, a monochromatic wave has been chosen as a seed for the instability.
As a result, no significant interaction between the shocks has been
observed. In a more realistic situation with a much longer precursor
and shocks of varying amplitudes, strong interactions between shocks
are to be expected \citep[\eg][]{gurbatov}. In particular, stronger
shocks overtake weaker ones and absorb them, so that the interaction
has a character of inelastic collisions. Time asymptotically, such
decaying shock turbulence is characterized by decreasing spatial density
of the shocks. 

In the context of the present study, we must consider driven, rather
than decaying, turbulence. A simple version of the driven Burgers
model that reveals shocks merging and creation of new shocks has been
considered by \cite{mkk95}. In the case of the driven turbulence
shocks not only merge but also new shocks are formed in-between. The
new shocks merge with their neighbors and so on. In a steady state,
some statistically stationary ensemble of shocks can be assumed with
certain average distance $L$ between the shocks and shock strength
characterized by an average Mach number $M$. For a systematic study
of particle propagation in such gas of shocks, the pdf's of Mach numbers
and distances between shocks are clearly required. As we mentioned
above, we make here only rough estimates of these quantities. In any
event, for the generation of statistically stationary shock ensemble,
the instability should act faster than the convection of a fluid element
across the precursor does, which requires $\gamma_{D}L_{p}/U_{1}\gg1$.
The latter condition can be transformed to the following 

\begin{equation}
\frac{P_{c}}{\rho U_{1}C_{s}}\gg1\label{eq:crit}\end{equation}
Note that the last condition does not contain the precursor scale
$L_{p}$ and can be also represented as $\epsilon M\gg1$, where $\epsilon=P_{c}/\rho_{1}U_{1}^{2}$.
With $P_{c}$ referring here to its subshock value, $\epsilon$ is
the acceleration efficiency and $M=U\left(z\right)/C_{s}$ is an acoustic
Mach number of the flow. Not surprisingly, the criterion $\epsilon M\gg1$
differs from the condition $\delta B^{2}/B_{0}^{2}\gg1$ only by a
factor of $M/M_{A}=1/\sqrt{\beta}$, as it follows from eq.(\ref{eq:maxdb}).
Indeed, the both instabilities share the same source of free energy,
the CR pressure gradient. 

Now that we can assume that the Drury instability has sufficient time
to fully develop to a nonlinear level we need to examine whether the
nonlinear stage of the wave steepening lasts longer than the linear
one. For the wave breaking time being shorter than the linear growth
it is sufficient to fulfill $\gamma_{D}<\Delta v/L$, where $\Delta v$
is the characteristic magnitude of the velocity jump across the shocks
and $L$ is the characteristic distance between them. The last condition
can be rewritten as 

\begin{equation}
\frac{L}{L_{p}}<\frac{\Delta v}{U_{1}}\frac{1}{\epsilon M}\label{eq:LLp1}\end{equation}
Under these circumstances, in a steady state shock merging must be
equilibrated with their growth due to the instability, \ie $\gamma_{D}\sim\delta v/L$,
where $\delta v$ is the characteristic variation of the shock strength,
\ie the rms difference between the neighbor values of $\Delta v$.
The latter condition can be written as 

\begin{equation}
\frac{L}{L_{p}}\sim\frac{\delta v}{U_{1}}\frac{1}{\epsilon M}\label{eq:LLp2}\end{equation}
Among the three parameters entering the last expression, $\delta v/U_{1}$,
$\epsilon$ and $M$, only the local Mach number $M\left(z\right)$
needs special attention. Indeed, the far upstream $M$ is considered
to be prescribed and may be very large but because of the multiple
shock formation in the CRP, strong heating should occur and drive
this parameter down considerably. 

Let us estimate the local Mach number $M$. There are two major factors
determining this parameter. After a fluid element passes through a
shock in the shocktrain it gets heated but it cools adiabatically
before it crosses the next shock. For this rough estimate we can assume
that all shocks are of the same compression ratio and the overall
flow pattern is $L$-periodic. For this order of magnitude type of
estimate, we neglect the gradual variation of the flow at the scale
$L_{p}>L$. We thus assume that the flow speed in front of each shock
is $v_{1}$ and it is equal to $v_{2}=v_{1}/r$ behind it, where $r$
is the shock compression ratio. Similarly, the densities are related
through $\rho_{2}=r\rho_{1}$. We estimate the shock heating from
the standard Rankine-Hugoniot relations. Let us denote the gas pressure
in front of a shock in the shocktrain as $P_{g1}$. Then, the gas
pressure behind it is

\[
P_{g2}=\frac{2\gamma M^{2}-\gamma+1}{\gamma+1}P_{g1}\]
 The subsequent decompression phase decreases the pressure so that
it becomes equal to $P_{g1}^{\prime}=r^{-\gamma}P_{g2}$ in front
of the next shock. Combining this with the above equation, we obtain

\[
\frac{P_{g1}^{\prime}}{P_{g1}}=\left(\frac{\gamma-1+2/M^{2}}{\gamma+1}\right)^{\gamma}\frac{2\gamma M^{2}-\gamma+1}{\gamma+1}\]
Therefore the total pressure change after passing trough the shock
and the smooth flow behind it is zero only for $M=1.$ However, the
pressure does not really change for a broad range of Mach numbers
around $M=1$ (where the r.h.s. of the last expression is close to
unity). This result means that $M$ cannot be large, since it would
lead to strong plasma heating driving its temperature to a state with
$M\simeq1$. However, in this situation the calculation of heating
using the Rankine-Hugoniot relations is not correct, since shocks
are likely to be sub-critical. Note that in our estimates, we also
ignored the increase of magnetic energy behind the shocks. Since the
above estimates show that shocks cannot be highly supercritical also,
we conclude that a reasonable estimate for $M$ to substitute in eq.(\ref{eq:LLp2})
is a critical value of the Mach number \ie $M\simeq M_{*}$ which
is of an order of a few ($M_{*}\simeq3,$ for example) depending on
plasma $\beta$ and $\vartheta_{nB}$- the angle between the shock
normal and the magnetic field \citep[see][for a review
of collisionless shock physics]{sagd66,pap}.

The efficiency $\epsilon$ may then be calculated once the heating
rate is known \citep[see, \eg][]{mdru01}). As for the parameter $\delta v/U_{1}$,
the first obvious constraint is $\delta v<\Delta v$ which obviously
verifies the consistency of eqs.(\ref{eq:LLp2}) and (\ref{eq:LLp1}).
In fact, based on the studies of the driven Burgers turbulence \citep{goto},
we can assume that the shock strength variance $\delta v$ and its
mean $\Delta v$ are related by%
\footnote{This rough estimate should be taken with caution. Both quantities
are really distributions, often with a power-law pdf, and may not
even posses finite means.%
} $\delta v\la\Delta v$. At the same time $\Delta v$ may be assumed
to be not so much weaker than the subshock itself \citep{drufal,kjr92}.
The subshock strength can be calculated from nonlinear acceleration
theory \citep{m97a,blasiSemAn,blasi05}, along with the acceleration
efficiency. Given the uncertainty of the heating rate one can estimate
$\epsilon\sim1/5-1/3$, so that the estimate $L/L_{p}\sim1/5$ does
not seem to be totally unreasonable. 

As it follows from the previous section, the maximum particle momentum
can be estimated from the relation $r_{g}\left(p_{max}\right)\sim L$.
Bearing in mind that $L_{p}\simeq r_{g}\left(p_{*}\right)c/U_{1}$,
the maximum momentum $p_{max}$ can be estimated as

\[
\frac{p_{max}}{p_{*}}\simeq\frac{c}{U_{1}}\frac{L}{L_{p}}\]
Since $L/L_{p}$, as we argued, can be not very small and $c/U_{1}$
is a large parameter, the suggested mechanism can produce a significant
additional acceleration beyond the break momentum at $p=p_{*}$. This
stage of acceleration lasts for approximately $\tau_{acc}\left(p_{max}\right)\sim\left[\kappa\left(p_{*}\right)/U_{1}^{2}\right]\ln\left(p_{max}/p_{*}\right)$.

\section{Particle spectrum between the break and maximum momentum\label{sec:Particle-spectrum-between}}

Up to now we have calculated the particle energy gain by their scattering
on unspecified scattering centers carried by the converging flow in
the CR precursor. Independently of that, we then considered the scattering
of particles on an ensemble of shocks. The latter process did not
lead to the energy gain, since no relative motion between the scatterers
was assumed. Recall that we considered the simple shock ensemble as
a magnetic structure traveling at a constant speed. Nevertheless,
each shock in the shocktrain may posses its own flow structure with
upstream and downstream regions and thus can in principle accelerate
particles via the standard DSA mechanism. However, as we pointed out
earlier, these shocks do not have sufficient resonant turbulence upstream
and downstream with $k<1/r_{g}\left(p_{*}\right)$. Therefore, there
is no suffiecient coupling of particles with momenta $p>p_{*}$ to
the converging flow around each such shock in order to gain momentum. 

In the situation we will consider further in this section, the momentum
gain results from the combination of particle scattering off the scattering
centers (shocks, localized magnetic structures) and gradual flow compression,
leading to the convergence of the centers. In fact this is similar
to what we have considered in Sec.\ref{sec:Acceleration-Mechanism}
at an elementary level. An essential difference is that the scattering
in momentum space is not homogeneous as the results of Sec. \ref{sec:Particle-dynamics}
suggest. Therefore, we relax the assumption, also made in Sec.\ref{sec:Acceleration-Mechanism},
that particle transport in momentum space is diffusion evenly covering
isoenergetic surfaces. This affects the formula for the momentum gain.

\subsection{Particle momentum gain}

To calculate the rate at which particles gain momentum let us write
down the equations of particle motion in the sub-shock reference frame
and apply them in the area ahead of the subshock where the velocity
of the flow is $\mathbf{U}\left(z,t\right)=U\left(z,t\right)\mathbf{e_{z}}$
(with $z$-axis pointing in the shock normal direction, $\mathbf{e_{z}}$) 

\begin{equation}
\frac{d\mathbf{p}}{dt}=\frac{e}{c}\left(\mathbf{v-U}\right)\times\mathbf{B}\label{eq:dpdt}\end{equation}
In this approximation, we have replaced the electric field by $\mathbf{E=-}c^{-1}\mathbf{U}\times\mathbf{B}$,
assuming that due to the perfect conductivity the electric field vanishes
in the local plasma frame. In contrast to Sec. \ref{sec:Particle-dynamics},
we do not specify the distance between the scattering centers here,
so it is convenient to normalize the length to $c/\omega_{ci}$, time
to $\omega_{ci}^{-1}$, $p$ to $mc$ while $v$ and $U\left(z\right)$
to $c$. 

Introducing the cosine of the particle pitch angle to the shock normal,
$\mu$, as well as the particle gyrophase $\phi$, as in Sec.\ref{sec:Particle-transport},
it is useful to rewrite the equations of motion given by eq.(\ref{eq:dpdt})
as 

\begin{eqnarray}
\dot{p} & = & \sqrt{1-\mu^{2}}U\Im\left(be^{-i\phi}\right)\label{eq:pdot1}\\
\dot{\mu} & = & \frac{1-\mu U}{pU}\dot{p}\label{eq:mudot1}\\
\dot{\phi} & = & -\frac{1}{p}+\frac{\mu-U}{\sqrt{1-\mu^{2}}}\Re\left(be^{-i\phi}\right)\label{eq:fidot1}\\
\dot{z} & = & \mu\label{eq:zdot2}\end{eqnarray}
We have introduced the following complex variable instead of the transverse
component of the magnetic field by $b=\left(B_{x}+iB_{y}\right)/B_{0}$.
The above equations are written in the shock frame. Now we need to
transform particle momentum $p$ to the local plasma frame, \ie the
frame moving with respect to the shock with the velocity $U\left(z,t\right)$
since it is this momentum the convection-diffusion equation {[}eq.(\ref{eq:difconv}){]}
refers to. This transformation can be written in the approximation
$U\ll1$ as $p^{\prime}=\left(1-U\mu\right)p$. Now we differentiate
the both sides of the last formula with respect to time, taking into
account that to this order of approximation $dU/dt\approx\dot{z}\partial U/\partial z$.
Making also use of eqs.(\ref{eq:mudot1}) and (\ref{eq:zdot2}), for
the acceleration rate in the local plasma frame we obtain

\begin{equation}
\dot{p}^{\prime}=-\mu^{2}\frac{\partial U}{\partial z}p^{\prime}\label{eq:pdot2}\end{equation}
 Recall that the precursor scale height $L_{p}$ that determines $\partial U/\partial z$
should be calculated using the spectral break momentum at $p_{*}$,
\ie $L_{p}=\kappa\left(p_{*}\right)/U_{1}$. We also note that the
acceleration rate in eq.(\ref{eq:pdot2}) does not contain the dynamical
variable $\phi$, and is actually almost independent of $z$. This
is because $U$, to a good approximation, can be considered as linear
function of $z$ in a significant part of the shock precursor (in
the case of the Bohm diffusion). As we mentioned earlier, due to the
large gyroradius, we can neglect any short scale variation of $U$.
This includes the structure of individual shocks in the shocktrain,
so that only the gradual variation of $U$ across the CRP is important.
Also, we are really interested  in averaged value of the acceleration
rate $\dot{p}^{\prime}/p^{\prime}$. For example, if the particle
transport in pitch angle is a small step diffusion, $\mu$ can be
regarded as random variable evenly distributed over the interval $\left(-1,1\right)$.
In this simple case, the average $\overline{\mu^{2}}=\frac{1}{3}$
and we recover the standard acceleration rate given by eq.(\ref{eq:pdot}).
In the case of structured phase space considered earlier in Sec. \ref{sec:Particle-dynamics},
one can assume that during most of an acceleration cycle, a particle
is bound to the island having $\overline{\mu}=0$. If $\overline{\mu}\neq0$,
the particle performs the Levy flights, which we discuss later. In
the case of small islands (\ie with the width $\mu_{0}\ll1$), we
can replace $\overline{\mu^{2}}$ by $\overline{\mu}^{2}$(for $\overline{\mu}\neq0$)
and by $\mu_{0}^{2}/3$ (for $\overline{\mu}=0$).

The situation becomes somewhat more complicated in the case of an
oblique shock, where the phase space shown, for example in Fig. \ref{sec:Particle-dynamics},
implies that the cosine of the pitch angle $\tilde{\mu}$ to be measured
with respect to the average magnetic field and not with respect to
the shock normal as in eq.(\ref{eq:pdot2}). The transformation to
the angles $\tilde{\mu}$ and $\tilde{\phi}$, \ie to the reference
frame having $z$-axis aligned with the averaged magnetic field, is
given by 

\begin{equation}
\mu=\tilde{\mu}\cos\vartheta-\sqrt{1-\tilde{\mu}^{2}}\sin\vartheta\sin\tilde{\phi}\label{eq:transf}\end{equation}
 where $\vartheta=\tan^{-1}\overline{B}_{y}$ (see Sec. \ref{sec:Particle-dynamics}).
Therefore, the acceleration rate given by eq.(\ref{eq:pdot2}) does
include the gyrophase $\tilde{\phi}$. However, the dynamics on the
$\left(\tilde{\phi},\tilde{\mu}\right)$ plane in the case of trapping
we are interested here and which is exemplified in Fig. \ref{cap:Chaos},
is mostly rotation around the origin, so that one can expect that
$\overline{\tilde{\mu}\sin\tilde{\phi}}\approx0$. Thus the acceleration
rate can be represented as 

\begin{equation}
\frac{\dot{p}}{p}\approx-K\frac{\partial U}{\partial z}\label{eq:accrate}\end{equation}
where 

\begin{equation}
K=\left[\frac{1}{3}\mu_{0}^{2}\cos^{2}\vartheta+\frac{1}{2}\left(1-\frac{1}{3}\mu_{0}^{2}\right)\sin^{2}\vartheta\right]\label{eq:Kdef}\end{equation}
and $\mu_{0}$ denotes the width of the island in which the particle
orbit is trapped. We also omitted the primes in momentum $p$. 

Using this formula it is easy to calculate particle momentum gain
assuming that the particle is trapped by the flow at the distance
$z$ from the subshock in the CRP where the local flow speed is $U\left(z\right)$
and is convected with the flow to the subshock where the flow speed
is $U_{0}$. Suppose that at the moment $t$ and coordinate $z$ the
particle has the momentum $p_{0}$. From eq.(\ref{eq:accrate}) we
have for the particle momentum at $z=0$ where $U=U_{0}$

\begin{equation}
p=p_{0}\exp\left[K\int_{t}^{t_{0}}\frac{\partial U}{\partial z}dt\right]=p_{0}\exp\left[K\ln\frac{U\left(z\right)}{U_{0}}\right]=p_{0}\left(\frac{U}{U_{0}}\right)^{K}\label{eq:momgain}\end{equation}
Note that in the conventional diffusive acceleration, where particles
ergodically cover the entire isoenergetic surface, ie $\mu_{0}=1$,
the acceleration constant $K=\frac{1}{3}$ (independent of $\vartheta$)
and the last result signifies the effect of adiabatic compression
of the CR gas by the converging flow. On the other hand, if the particle
dynamics is such that $\mu_{0}\approx0$, $\vartheta\approx\pi/2$,
which corresponds to two-dimensional compression across magnetic field,
one obtains $K\approx\frac{1}{2}$. In strong CR modified shocks,
the shock precompression $R=U_{1}/U_{0}$ scales as $M^{3/4}$ \citep{kaz86,byk96}.
As it was shown by \cite{m97a}, the latter scaling is only valid
for $M<M_{*}\equiv\left(\nu_{\textrm{inj}}p_{*}/p_{inj}\right)^{4/3}$
, and the shock pre-compression $R$ saturates at $R\sim\nu_{inj}p_{*}/p_{inj}$
when $M>M_{*}$  \cite[see also][]{blasi05}. Here the injection parameter
$\nu_{inj}\simeq\left(cp_{inj}/mU_{1}^{2}\right)n_{CR}/n_{1}$ and
$p_{inj}$ is the injection momentum while $n_{CR}$ and $n_{1}$
are the number density of accelerated particles and that of the background
plasma far upstream, respectively. In both Mach number ranges the
shock precompression can be quite significant unless the gas heating
in the CRP is strong. At the same time, strong reduction of the shock
precompression by plasma heating would diminish the heating itself
by weakening the acoustic (Drury's) instability of the CRP. The situation
may settle at some critical level where a moderate but still quite
significant precompression is accompanied by the CRP heating caused
by the acoustic instability \citep{mdv00,mdru01}. Further interesting
analysis of the nonlinear shock acceleration and its bifurcation has
been published recently by \cite{blasi05}.

\subsection{Acceleration model: details\label{sub:Acceleration-model:-details}}

\begin{figure}
\plotone{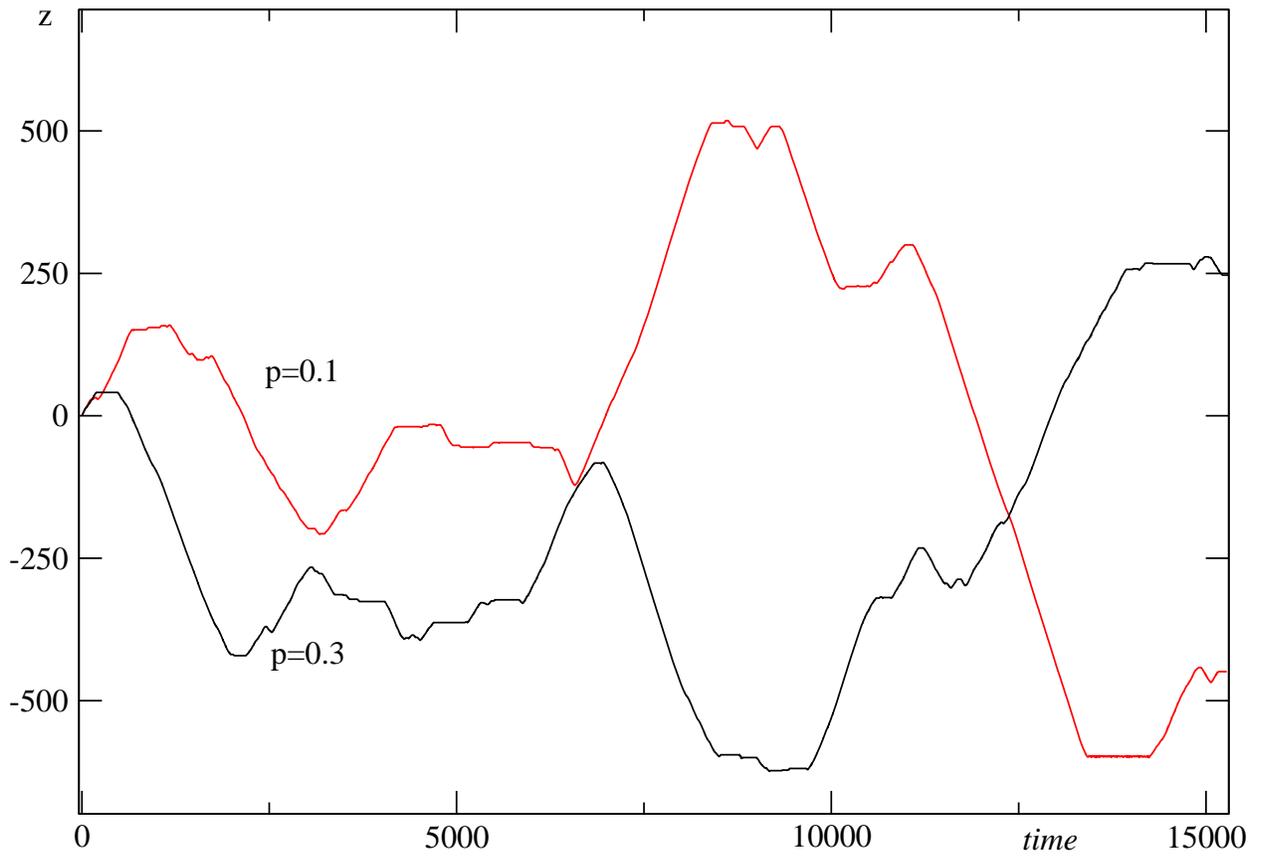}

\caption{The same as Fig.\ref{cap:zOft} but for smaller particle momenta,
$p=0.1$ and $p=0.3$.\label{cap:zOft2}}
\end{figure}

The picture that emerges from the above study of particle dynamics
in the modified shock precursor can be described as follows. There
are three groups of particles. One group is made up by particles that
are locally trapped in the plasma flow by a structure in the shocktrain.
They are convected with the flow, and either do not propagate at all
or propagate very slowly with respect to the flow. They are clearly
seen in \eg Fig.\ref{cap:zOft2} as the quasi-horizontal portions
of the stochastic trajectory of a single particle. The remaining two
groups of particles are particles performing Levy flights, \ie those
propagating ballisticaly in positive and negative direction at the
speed $U_{+}$ and $U_{-}$, respectively. These are characterized
by steep portions of the trajectory in Fig.\ref{cap:zOft2}, with
the positive and negative slopes, respectively. Let us denote the
phase space density of the above three groups of particles by $F_{0}\left(t,z,p\right)$
and $F_{\pm}\left(t,z,p\right)$. Furthermore, we introduce the probabilities
of transition in the unit of time by denoting by $\alpha_{\pm}$ the
probability of trapping of particles performing Levy flight in positive
and negative direction, respectively. We denote the rates of the reverse
transitions by $\beta_{\pm}$. Now we can write the balance of particles
in a phase space cell between $z_{1},\, p_{1}$ and $z_{2}=z_{1}+dz,\, p_{2}=p_{1}+dp$
as follows

\begin{eqnarray}
\frac{\partial}{\partial t}F_{0}\left(t,z,p\right)dzdp & = & \left(\left.F_{0}U\right|_{z_{1}}-\left.F_{0}U\right|_{z_{2}}\right)dp+\left(\left.\dot{p}F_{0}\right|_{p_{1}}-\left.\dot{p}F_{0}\right|_{p_{2}}\right)dz\label{eq:F0}\\
 & - & \left(\beta_{+}+\beta_{-}\right)F_{0}dzdp+\left(\alpha_{+}F_{+}+\alpha_{-}F_{-}\right)dzdp\end{eqnarray}
or after retaining only the linear terms in $dz$ and $dp$ and using
eq.(\ref{eq:accrate}) we obtain the following equation for the phase
space density of trapped particles $F_{0}$

\begin{equation}
\frac{\partial F_{0}}{\partial t}+\frac{\partial}{\partial z}UF_{0}-K\frac{\partial U}{\partial z}\frac{\partial}{\partial p}pF_{0}=-\left(\beta_{+}+\beta_{-}\right)F_{0}+\alpha_{+}F_{+}+\alpha_{-}F_{-}\label{eq:DCF0}\end{equation}
Here the constant $K$ is given by eq.(\ref{eq:Kdef}). Proceeding
in a similar manner we obtain two equations for $F_{\pm}$, \ie for
the phase space density of the particles in the Levy flight state

\begin{equation}
\frac{\partial F_{\pm}}{\partial t}+\frac{\partial}{\partial z}U_{\pm}F_{\pm}-K_{\pm}\frac{\partial U}{\partial z}\frac{\partial}{\partial p}pF_{\pm}=-\alpha_{\pm}F_{\pm}+\beta_{\pm}F_{0}\label{eq:Fpm}\end{equation}
In contrast to eq.(\ref{eq:DCF0}) the acceleration constants $K_{\pm}$
are defined by 

\begin{equation}
K_{\pm}=\left[\mu_{\pm}^{2}\cos^{2}\vartheta+\frac{1}{2}\left(1-\mu_{\pm}^{2}\right)\sin^{2}\vartheta\right]\label{eq:Kpm}\end{equation}
where $\mu_{\pm}$ are the averaged values of the cosines of the particles
pitch angles ($\mu_{\pm}=\overline{\tilde{\mu}}$) performing Levy flights
in the positive and negative direction, respectively, while $U_{\pm}=c\mu_{\pm}\cos\vartheta\gg U$
are their speed components along the shock normal, cf. eq.(\ref{eq:Kdef}).
Note that we ignore direct transitions between the opposite Levy flights
as comparatively rare events. This can be inferred from Fig.\ref{cap:zOft2},
for example.

Equations (\ref{eq:DCF0}) and (\ref{eq:Fpm}) form a closed system
that must be supplemented with the boundary conditions. Far upstream
from the shock, at $z=\infty$, it is natural to impose the following
boundary conditions

\begin{equation}
F_{0}\left(\infty\right)=F_{-}\left(\infty\right)=0\label{eq:BCinf}\end{equation}
which simply mean that accelerated particles can only leave the system
in that direction, but do not enter from it. Note, that $F_{+}\left(\infty\right)\neq0$
and should be determined from eqs.(\ref{eq:F0}) and (\ref{eq:Fpm}).
The second boundary condition, the one at the sub-shock location $z=0$
is somewhat more ambiguous. In the standard DSA scheme it is fixed
by the flux of particles convected with the flow downstream. It can
be expressed through the particle phase space density and the downstream
flow speed since the particle distribution is isotropic. In the case
of high momenta ($p>p_{*}$) considered here, particles are not bound
to the flow so strongly as to be able to make an isotropic distribution.
Rather, according to eqs.(\ref{eq:F0}) and (\ref{eq:Fpm}), particles
convected with the flow at the speed $U\left(z\right)$ are being
converted into particles propagating at high sped $U_{\pm}\sim c$
and can either leave the system upstream (unless they are re-trapped)
or can penetrate deeply downstream and reach the contact discontinuity.
The latter possibility for the high energy particles has been already
discussed in the literature \citep{ber96,blond}. The fact that the
contact discontinuity and the forward shock may indeed be closer to
each other due to the backreaction of accelerated particles on the
shock dynamics, as clearly follows from the nonlinear DSA, seems to
find impressive observational confirmation in \citep{warren}. The
strong magnetic field at the contact discontinuity may very well reflect
particles with $p>p_{*}$ and they can return to the shock. In a simple
form this requirement can be obviously formulated mathematically setting
the total particle flux through $z=0$ equal to zero:

\begin{equation}
F_{0}\left(0\right)U_{0}+F_{-}\left(0\right)U_{-}+F_{+}\left(0\right)U_{+}=0\label{eq:BCzero}\end{equation}
In other words, the negative part of the total particle flux, represented
by the first two terms ($U_{0,-}<0$) is reflected by the strong magnetic
turbulence downstream and ultimately reappears at the shock in the
form of the untrapped particles propagating upstream at the speed
$U_{+}>0$. The strong turbulence downstream may be formed by the
magnetic perturbation convected and compressively amplified from the
upstream media and by the Raleigh-Taylor instability of the contact
discontinuity \citep{JJ}.

The way particle momentum enter eqs.(\ref{eq:DCF0}) and (\ref{eq:Fpm})
as well as the homogeneity of the boundary conditions given by eqs.(\ref{eq:BCinf})
and (\ref{eq:BCzero}), suggest the power-law solution:

\begin{equation}
F_{0,\pm}\propto p^{-q}\label{eq:powl}\end{equation}
Assuming for simplicity $U_{\pm}=\const$, (which is close to the
truth, as \eg Fig.\ref{cap:zOft2} suggests) we can rewrite these
equations as follows

\begin{eqnarray}
\frac{d}{dz}UF_{0} & = & \gamma_{0}F_{0}+\alpha_{+}F_{+}+\alpha_{-}F_{-}\label{eq:F0red}\\
U_{\pm}\frac{dF_{\pm}}{dz} & = & \gamma_{\pm}F_{\pm}+\beta_{\pm}F_{0}\label{eq:Fpmred}\end{eqnarray}
Here we introduce the following notations

\begin{eqnarray*}
\gamma_{0} & = & -K\left(q-1\right)\frac{\partial U}{\partial z}-\beta\\
\gamma_{\pm} & = & -K_{\pm}\left(q-1\right)\frac{\partial U}{\partial z}-\alpha_{\pm}\end{eqnarray*}
where

\begin{equation}
\beta=\beta_{+}+\beta_{-}\label{eq:beta}\end{equation}
From eqs.(\ref{eq:F0red}) and (\ref{eq:BCzero}), we have

\begin{equation}
F_{0}\left(z\right)=-\frac{1}{U}\int_{z}^{\infty}\exp\left(-\int_{z}^{z^{\prime}}\frac{\gamma_{0}}{U}dz^{\prime\prime}\right)\left[\alpha_{+}F_{+}\left(z^{\prime}\right)+\alpha_{-}F_{-}\left(z^{\prime}\right)\right]dz^{\prime}\label{eq:F0sol}\end{equation}
To zeroth approximation in $U/U_{\pm}\ll1$, from eq.(\ref{eq:Fpmred})
we have $F_{+}\approx\const$ and $F_{-}\approx\const$, while from
the boundary condition given by eq.(\ref{eq:BCinf}) and (\ref{eq:BCzero})
we obtain

\[
F_{+}=-\frac{U_{0}}{U_{+}}F_{0}\left(0\right)\]
and $F_{-}\approx0$. Upon substitution of $F_{\pm}$ into eq.(\ref{eq:F0sol})
and taking it at $z=0$ we finally obtain the following equation for
for the spectral index $q$:

\begin{equation}
\int_{0}^{\infty}\left[\frac{U\left(z\right)}{U_{0}}\right]^{K\left(q-1\right)}\exp\left(\int_{0}^{z}\frac{\beta}{U}dz^{\prime}\right)\alpha_{+}\left(z\right)\frac{dz}{U_{+}}=1\label{eq:lamdaeq}\end{equation}
In fact, the last formula allows a very simple physical interpretation,
making it essentially equivalent to the well known general \cite{fermi49}
result. Indeed, Fermi showed that if a group of particles undergoes
continuous acceleration at the rate $\tau_{acc}^{-1}$ and have certain
probability to escape that can be characterized by the escape time
$\tau_{esc}$, then the energy (momentum distribution normalized above
as $F_{0}\left(p\right)$) has the following spectral index

\begin{equation}
q_{F}=1+\frac{\tau_{acc}}{\tau_{esc}}\label{eq:qF}\end{equation}
At least for some short period of time, by definition of these time
scales, one can state that the momentum gain $p\left(t\right)/p\left(0\right)\equiv p/p_{0}=\exp\left(t/\tau_{acc}\right)$.
The probability to stay in the accelerator if escape is a Poisson
process, $\mathcal{P}_{conf}=\exp\left(-t/\tau_{esc}\right)$. Then
eq.(\ref{eq:qF}) can be recast as  \citep[cf.][]{bell78a,be87,acht01}

\begin{equation}
q_{F}=1+\frac{\ln\left(1/\mathcal{P}_{conf}\right)}{\ln\left(p/p_{0}\right)}\label{eq:qF2}\end{equation}
Note that $t$ can be regarded here as the duration of an acceleration
cycle and it does not enter the spectral index explicitly. It should
be also clear that determination of $\mathcal{P}_{conf}$ and the
momentum gain (or in other words, $\tau_{esc}$ and $\tau_{acc}$)
requires in our model more information about particle transport, in
particular, the particle trapping and untrapping rates $\alpha$ and
$\beta$. 

To demonstrate that the index $q$ in our formula (\ref{eq:lamdaeq})
must be interpreted in the same way as in the classical Fermi problem,
let us analyze various factors in the integrand in eq.(\ref{eq:lamdaeq})
separately. Assume that a particle enters an acceleration cycle at
$z=0$ as a Levy flight in a positive direction with a speed $U_{+}$
against the plasma flow of speed $U\left(z\right)$ in the shock precursor,
located in a half-space $z>0$. During its flight, the particle has
a possibility to be trapped in the flow that is characterized by the
rate $\alpha_{+}$. In other words, during time $dt$ the probability
to be trapped between $z$ and $z+dz$, equals $\alpha_{+}\left(t\right)dt=\alpha_{+}\left(z\right)dz/U_{+}$.
We see this probability density in the integral given by eq.(\ref{eq:lamdaeq}).
Assume that the trapping event actually occurred at some $z$. Then
the particle is convected (with the flow) back to the subshock at
the speed $U\left(z\right)$. During this convection it has a certain
probability to escape, which is characterized by the rate $\beta$.
This means that if the particle has a probability of being in the
flow $P\left(t\right)$, then the probability to remain there at $t+dt$
is ${\mathcal{P}}\left(t+dt\right)=\mathcal{P}\left(t\right)\left(1-\beta dt\right)$,
or if the particle is certainly in the flow at $t=0$ (just trapped
there, for example) then $\mathcal{P}\left(t\right)=\exp\left(-\int_{0}^{t}\beta dt\right)$.
Noting that $dt=dz/U$, we can write the probability of returning
with the flow from $z$ to $z=0$ as

\[
\exp\left(-\int_{z}^{0}\beta dz/U\right)\]
Combining with the above probability of being trapped between $z$
and $z+dz$, we can express the probability of completing an acceleration
cycle of the length between $z$ and $z+dz$ as%
\footnote{Particle behavior during a Levy flight is non-Markovian, since it
bears elements of deterministic motion (this is also true for the
trapped particles). However, the joint probability of two such events
is multiplicative since they evolve independent of each other. %
} 

\[
d\mathcal{P}_{cycl}\left(z\right)=\exp\left(\int_{0}^{z}\frac{\beta}{U}dz^{\prime}\right)\alpha_{+}\left(z\right)\frac{dz}{U_{+}}\]
The total probability of completing one such cycle of any possible
length $0<z<\infty$ is 

\[
\mathcal{P}_{cycl}=\int_{0}^{\infty}\frac{d\mathcal{P}_{cycl}}{dz}dz\]
We can now rewrite eq.(\ref{eq:lamdaeq}) as follows

\begin{equation}
\int_{0}^{\infty}\left[\frac{U\left(z\right)}{U_{0}}\right]^{K\left(q-1\right)}\frac{d\mathcal{P}_{cycl}}{dz}dz=1\label{eq:lamdaeq2}\end{equation}
Here the quantity $G\left(z\right)=\left[U\left(z\right)/U_{0}\right]^{K}$
is equal to a particle momentum gain $p_{f}/p_{i}$ after it is convected
with the flow from $z$ to the origin, eq.(\ref{eq:momgain}). Using
the mean value theorem, eq.(\ref{eq:lamdaeq2}) can be represented
as

\[
\left\langle \frac{p_{f}}{p_{i}}\right\rangle ^{q-1}\mathcal{P}_{cycl}=1\]
where $\left\langle p_{f}/p_{i}\right\rangle =G\left(\overline{z}\right)$
is the mean momentum gain per cycle and $\mathcal{P}_{cycl}$ is,
as stated above, the probability to complete this acceleration cycle.
From the last formula we obviously have

\begin{equation}
q=1+\frac{\ln\left(1/\mathcal{P}_{cycl}\right)}{\ln\left\langle p_{f}/p_{i}\right\rangle }\label{eq:qfin}\end{equation}
which is formally equivalent to the classical Fermi result. 

\begin{figure}
\plotone{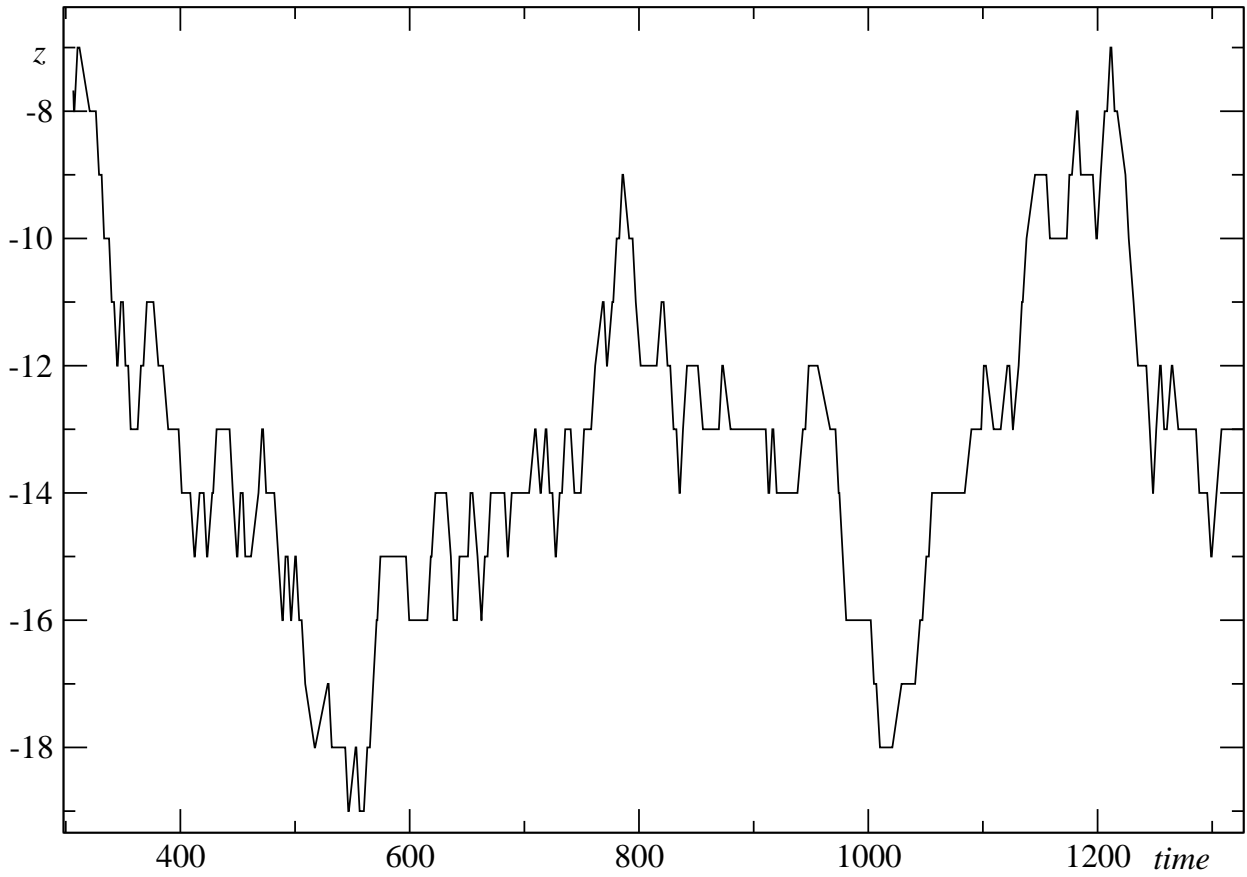}

\caption{Stochastic particle trajectory represented in the same format as
in Fig.\ref{cap:zOft2} but for $p=0.3$, $\overline{B}_{y}=1$ and $\tilde{B}_{y}=2$.\label{cap:zOft3}}
\end{figure}

The quantities $\alpha_{+}$ and $\beta$, which the index $q$ depends
upon, can be inferred from the particle stochastic trajectories shown,
\eg  in Figs. \ref{cap:zOft} and \ref{cap:zOft2}. Indeed the probability
for a particle to be in the trapped state can be written as $\tau_{tr}/\left(\tau_{tr}+\tau_{L}^{+}+\tau_{L}^{-}\right)$
where the $\tau_{tr}$ and $\tau_{L}$ are the average trapping and
Levy flight times that can be calculated from the particle trajectory.
Then, the particle trapping rate $\alpha_{+}$ can be written as 

\[
\alpha_{+}\approx\frac{\tau_{tr}}{\left(\tau_{tr}+\tau_{L}^{+}+\tau_{L}^{-}\right)}\frac{1}{\tau_{L}^{+}}\]
similarly

\[
\beta_{\pm}\approx\frac{\tau_{L}^{\pm}}{\left(\tau_{tr}+\tau_{L}^{+}+\tau_{L}^{-}\right)}\frac{1}{\tau_{tr}}\]
For the case of sufficiently strong shocks in the shock precursor,
the trapping process is quite efficient, so we can assume that

\begin{equation}
\int_{0}^{L_{p}}\frac{\beta}{U}dz\ll1.\label{eq:betacond}\end{equation}
Here the shock precursor length $L_{p}$ can be related to the trapping
probability as

\[
L_{p}\sim\int_{0}^{\infty}\alpha_{+}\left(z\right)zdz\diagup\int_{0}^{\infty}\alpha_{+}\left(z\right)dz.\]
The requirement for the inequality in eq.(\ref{eq:betacond}) is then

\begin{equation}
\frac{\tau_{p}\tau_{L}}{\tau_{tr}^{2}}\ll1\label{eq:betcond2}\end{equation}
where the precursor crossing time is

\[
\tau_{p}=\int_{0}^{L_{p}}\frac{dz}{U}\]
An example of a particle trajectory with such long trapping times
is shown in Fig.\ref{cap:zOft3}. Note that when a particle is bouncing
between 2-3 neighboring shocks it can also be considered as {}``trapped''
in the flow, as is the case for a particle stuck to a single shock.
These kind of events are clearly seen in Fig.\ref{cap:zOft3}.

\section{Summary and Conclusions\label{sec:Conclusions}}

The principal conclusion of this paper is that particle acceleration
inside the cosmic ray shock precursor may very well be faster and
proceed to higher energies during SNR shock evolution than the conventional
DSA theory suggests. The requirement for such enhanced acceleration
is some limitation on the momentum of particles that contribute the
most to the CR pressure in order to prevent further inflation of the
shock precursor. This can be achieved by spatial limitations of momentum
growth, as discussed in detail by \eg \cite{ber96}, by the wave
compression and blue-shifting of the wave spectrum in the CRP \citep{mdj02}
and by the change of acceleration regime after the end of the free
expansion phase \citep{DruryKnee}. As a result of these limitations,
a break on the particle spectrum is formed and maintained at $p=p_{*}$
beyond which particles do not contribute to the pressure significantly,
but are still accelerated at the same rate as particles with momenta
$p<p_{*}$. 

The fundamental acceleration mechanism is essentially identical to
the nonlinear DSA. \emph{The difference from the linear DSA is that
frame switching which results in the energy gain occurs predominatingly
between the scatterers convected with the gradually converging upstream
flow in the precursor, and not between the upstream and downstream
scattering centers, as usually assumed.} The maximum momentum is estimated
to be

\begin{equation}
p_{max}\sim\frac{c}{U_{1}}\frac{L}{L_{p}}p_{*}\label{eq:pmaxfin}\end{equation}
with the maximum value of $L/L_{p}\la1/5$. Therefore, this mechanism
provides up to $\sim.2\left(c/U_{1}\right)$ enhancement to the maximum
momentum, based on the standard spatial constraints \citep[\eg][]{ber96}.
The spectral index between $p_{*}$ and $p_{max}$ depends on both
the ratio of the particle trapping to flight time $\tau_{tr}/\tau_{L}$
in the upstream turbulent medium, and on the shock precompression
$R$. This is in a deep contrast with the spectrum below $p_{*}$,
which tends to have a form quasi-independent of $R$  \citep[\eg][]{mdru01}.
The acceleration time is 

\begin{equation}
\tau_{acc}\left(p_{max}\right)\sim\tau_{NL}\left(p_{*}\right)\ln\frac{p_{max}}{p_{*}}\label{eq:tauaccfin}\end{equation}
 where $\tau_{NL}\left(p\right)\simeq4\kappa\left(p\right)/U_{1}^{2}$
is the nonlinear acceleration time which only slightly differs from
the upstream contribution to the standard linear acceleration time
\citep{mdru01}.

The overall temporal development of the acceleration process can be
described as follows. It starts as the standard DSA mechanism from
some slightly suprathermal momentum $p_{inj}$ and proceeds at the
rate $\tau_{acc}^{-1}\left(p_{max}\right)\sim U_{1}^{2}/\kappa\left(p_{max}\right)$,
where $\kappa\left(p\right)\sim cr_{g}\left(p\right)$ and $r_{g}\left(p_{max}\right)k_{min}\sim1$.
The minimum wave number $k_{min}$ decreases with growing $p_{max}$$\left(t\right)$,
since the MHD waves confining particles with momenta $p_{inj}<p<p_{max}$
to the shock are resonantly generated by the accelerated particles
themselves. After $p_{max}$ reaches a certain value \citep[that can be
calculated analytically, \eg][]{mdru01} and is clearly seen in
numerical time dependent solutions \cite[\eg][]{byk96} a CRP forms
and the acceleration then proceeds mostly in the CRP. The acceleration
rate, however, behaves with time approximately as it does at the previous
(linear) stage since $dU/dz$ decreases with the growing width of
the CRP, $L_{p}\sim\kappa\left(p_{max}\right)/U_{1}$. Note that $dU/dz$
sets the acceleration rate. The next acceleration boosting transition
occurs when $L_{p}$ reaches one of its above mentioned {}``natural''
limits, such as the accelerator size (a significant fraction of an
SNR radius, for example, \ie  $\kappa\left(p_{max}\right)/U_{1}\sim R_{SNR}$).
Assume $p_{max}\simeq p_{*}$ at this point. Then particles with momenta
$p>p_{*}$ cannot be confined to the shock diffusively, since their
diffusion length $L_{dif}\left(p\right)=\kappa\left(p\right)/U_{1}>L_{p}\simeq\kappa\left(p_{*}\right)\textrm{/U}_{1}$.
Fortunately, by this time instabilities of the CRP produce an ensemble
of shocks or similar nonlinear magnetic structures. These ensembles
of structures turn out to be capable of confining particles with $p>p_{*}$,
since the mean distance between the structures (which is the scale
which replaces the wavelength of particle-confining turbulence in
the quasi-linear picture) is much longer than the particle gyroradius,
$L\gg r_{g}\left(p_{*}\right)$. Particle confinement is, however,
selective in terms of particle location in phase space. This is similar
to, but more general than, the particle confinement in magnetic traps
where only particles having pitch angles satisfying $\left|\mu\right|<\sqrt{\Delta B/B}$
are trapped while others leave the system freely. The price to be
paid for this is a spectrum at $p>p_{*}$ which is even steeper than
$p^{-4}$. This is in fact more a blessing than a curse since these
particles do not contribute to the CR pressure significantly and $L_{p}$
does not grow maintaining thus the acceleration rate at the same level
$U_{1}^{2}/\kappa\left(p_{*}\right)$ for all particles with momenta
$p_{*}<p<p_{max}$. 

The total acceleration time up to the maximum momentum $p_{max}$
(given by eq.{[}\ref{eq:tauaccfin}{]}) is thus only logarithmically
larger than the acceleration time to reach the spectral break (knee)
at $p\simeq p_{*}$ and the maximum momentum itself is also pinned
to the break momentum $p_{*}$ through eq.(\ref{eq:pmaxfin}). The
break momentum is simply the maximum momentum in the standard DSA
scheme, whatever physical process stops it from growing further. The
realization of the above scenario, which obviously produces a significant
acceleration enhancement in terms of both the maximum energy and acceleration
time, depends crucially on the presence of the ensemble of shocks
or similar scattering magnetic structures in the CRP. Their existence
in various (even CR unmodified) collisionless shocks is perhaps beyond
any reasonable doubt, as we discussed in Sec.\ref{sec:Particle-dynamics},
but the real challenge is to calculate the statistical characteristics
of their spacing $L$ and amplitude parameter $\Delta v$. A simple
approach that we pursued in Sec.\ref{sec:Estimate-of-the maxmom}
leads to the conservative estimate of $L/L_{p}\sim1/10$, so that
taking $U_{1}/c\sim10^{3}$, the maximum momentum exceeds its standard
value by a factor of $100$, eq.(\ref{eq:pmaxfin}). 

At this point it is worth while to recapitulate the major differences
between our approach and other recent models discussed in the Introduction
section which are also aimed at the enhancement of acceleration efficiency.
They primarily seek to increase the turbulent magnetic energy by tapping
the free energy of (already) accelerated particles. This would result
in decreasing the particle dwelling time (more frequent shock crossing,
and thus faster acceleration) and, concomitantly, the particle diffusion
length (better confinement, higher maximum energy). Our approach is
based on the appreciation of

(i) in a nonlinear regime, the acceleration time is reduced because
of a steeper velocity profile in the smooth part of the shock transition
(where acceleration mostly occurs). The steep gradient is maintained
due to the formation of a break on the particle spectrum. Without
the break, the gradient would flatten with growing particle energy
slowing down the acceleration. Apart from this gradient, the acceleration
rate does not depend on the m.f.p. explicitely, as long as the latter
is smaller than the shock transition. 

(ii) particle confinement (maximum energy) is enhanced by increasing
the wavelength (shock spacing in a shocktrain inside the CRP, for
example) of turbulence which is in a distinctively strong regime.
Departure from the weak turbulence is marked by modification of particle
confinement in the CRP. This steepens the momentum spectrum beyond
the spectral break and ensures its existence, as required in (i). 

It remains unclear, whether the realization of the first (\ie strong
magnetic field amplification) hypothesis would preclude or otherwise
strongly influence the acceleration scenario suggested in this paper.
It would almost certainly not, if the magnetic energy increases predominantly
at longest scales as suggested by \cite{dm04}. This should merely
scale the estimates provided in this paper by renormalising such basic
quantities as $L_{p}$ and $\tau_{NL}$, thus making acceleration
improvement even stronger. One disadvantage of acceleration enhancement
with significant field amplification, however, is that the increased
magnetic field inevitably leads to increased radiative losses with
a reduction of the particle maximum energy as an implication. This
important acceleration constraint has been studied in detail by \cite{AharDer},
including applications to many popular acceleration sites. Naturally,
the most prominent impact is expected on the acceleration of the extremely
high energy CRs (EHECR, $\ga10^{20}$eV). Detailed models of such
acceleration have been proposed by \cite{mini,kang05} and others. Clearly, the
acceleration improvement without significant increase of the magnetic
field potentially provides a significant edge on the field amplification
scenario in the general context of the problem of EHECR acceleration.

\acknowledgements{Support by NASA grant ATP03-0059-0034 and under U.S. DOE grant No.
FG03-88ER53275 is gratefully acknowledged }

\pagebreak


%

\end{document}